\documentclass[onecolumn]{aastex62}

\usepackage{times}

\usepackage{amsmath,amssymb,amstext}
\usepackage{tabularx}
\usepackage{longtable}
\usepackage{float}
\usepackage{natbib}

\newcommand{\kms}{km\,s$^{-1}$}

\begin{document}

\title{{\large{\bf First Detection of Plasmoids from Breakout Reconnection on the Sun}}}
\author{Pankaj Kumar}
\affiliation{Heliophysics Science Division, NASA Goddard Space Flight Center, Greenbelt, MD, 20771, USA}

\author{Judith T. Karpen}
\affiliation{Heliophysics Science Division, NASA Goddard Space Flight Center, Greenbelt, MD, 20771, USA}

\author{Spiro K. Antiochos}
\affiliation{Heliophysics Science Division, NASA Goddard Space Flight Center, Greenbelt, MD, 20771, USA}

\author{Peter F. Wyper}
\affiliation{Department of Mathematical Sciences, Durham University, Durham DH1 3LE, UK}

\author{C. Richard DeVore}
\affiliation{Heliophysics Science Division, NASA Goddard Space Flight Center, Greenbelt, MD, 20771, USA}

\email{pankaj.kumar@nasa.gov}

\begin{abstract}
 Transient collimated plasma ejections (jets) occur frequently throughout the solar corona, in active regions, quiet Sun, and coronal holes. Although magnetic reconnection is generally agreed to be the mechanism of energy release in jets, the factors that dictate the location and rate of reconnection remain unclear. Our previous studies demonstrated that the magnetic breakout model explains the triggering and evolution of most jets over a wide range of scales, through detailed comparisons between our numerical simulations and high-resolution observations. An alternative explanation, the resistive-kink model, invokes breakout reconnection without forming and explosively expelling a flux rope. Here we report direct observations of breakout reconnection and plasmoid formation during two jets in the fan-spine topology of an embedded bipole. For the first time, we observed the formation and evolution of multiple small plasmoids with bidirectional flows associated with fast reconnection in 3D breakout current sheets in the solar corona. The first narrow jet was launched by reconnection at the breakout current sheet originating at the deformed 3D null, without significant flare reconnection or a filament eruption. In contrast, the second jet and release of cool filament plasma were triggered by explosive breakout reconnection when the leading edge of the rising flux rope formed by flare reconnection beneath the filament encountered the preexisting breakout current sheet. These observations solidly support both reconnection-driven jet models: the resistive kink for the first jet, and the breakout model for the second explosive jet with a filament eruption.
\end{abstract}

\section{Introduction}
Coronal jets are collimated plasma ejections that occur repeatedly everywhere on the Sun and may contribute a significant amount of mass and energy to the corona and solar wind \citep{patsourakos2008,moore2010,raouafi2010,sterling2015,innes2016,raouafi2016}. Recurrent jets often are ejected from coronal bright points throughout their lifetimes of hours to days \citep{kumar2019}. The time scale and magnitude of the energy buildup are smaller, and the magnetic configurations are much simpler, in jet source regions than in the complex active regions that produce the most energetic large-scale eruptions (eruptive flares and coronal mass ejections (CMEs)). There are sufficient underlying commonalities in their magnetic structure and explosive dynamics, however, that jets provide an excellent opportunity to understand the energy buildup and release processes for the full range of solar eruptions.

Two important features of coronal-hole jets have emerged recently, due to the availability of high-resolution multiwavelength data. Most of these events appear to be associated with mini-filament eruptions \citep{sterling2015}, and many exhibit helical, untwisting motions \citep{patsourakos2008,innes2016}.  Magnetic reconnection is generally agreed to be the energy-release mechanism, but the location and timing of reconnection remain actively debated.  Flux emergence has been proposed as the driver of coronal jets, through reconnection between the preexisting field and the emerging flux systems \citep[e.g.,][]{shibata1994,moreno2013}. However, recent observations of jets do not support the flux emergence model \citep{kumar2018,kumar2019,McGlasson2019}. 

We have developed two reconnection-driven models for coronal jets. The most basic requirement for both is a multipolar photospheric magnetic field, the simplest manifestation of which is the embedded bipole, also known as a fan-spine topology \citep{priest1996}. Our resistive-kink model for coronal jets \citep{pariat2009,pariat2010, pariat2015, pariat2016, wyper2016a,wyper2016b, karpen2017} invokes common photospheric motions or emergence of sheared flux to add free energy to the magnetic field within the embedded bipole, in the form of a large-scale twist, without creating a flux rope. Expansion of this core field compresses and distorts the null, creating the breakout current sheet (BCS) and initiating slow reconnection there. Onset of an ideal tilting or kinking of the twisted flux impulsively launches fast reconnection between the core and external fields through the BCS, producing an Alfv\'enic helical jet but no flare reconnection or arcade formation. Our magnetic breakout model \citep{antiochos1998,antiochos1999,macneice2004,lynch2008,devore2008,karpen2012}, originally developed to explain fast coronal mass ejections (CMEs), has been proved to be a robust mechanism for producing solar eruptions on all scales, from jets with mini-filament eruptions to CMEs \citep{wyper2017,wyper2018}. This model invokes magnetic reconnection in two key locations to disrupt the force balance that maintains a strongly sheared, pre-eruptive configuration in the corona: breakout reconnection at a current sheet formed at a stressed null, and flare reconnection in the current sheet formed behind the rising sheared core field. Numerous observations exhibit features consistent with this comprehensive theoretical model for jets \citep{sterling2015,kumar2018,moore2018,panesar2018,kumar2019,li2019}. In fact, prior to the first jet discussed here, we had not found a definitive example of an observed resistive-kink jet. For purposes of this study, we will focus on the breakout reconnection that is common to both jet models.
 
Proof of flare reconnection is easily derived from numerous observations of flare arcades and ribbons \citep{forbes1996,fletcher2001}, and more recently from high-resolution observations of current sheets, inflows, outflows, footpoint dimmings, and plasmoids below the accompanying jet or CME \citep{takasao2012,reeves2015,kumar2018}.  Breakout reconnection is more difficult to detect than flare reconnection, because it occurs in the higher, more rarefied corona and usually is less energetic. Distinctive breakout signatures identified in simulations and observations include extreme ultraviolet (EUV) brightenings at the footpoints of the overlying separatrix/breakout current sheet, \citep{sterling2001,masson2009,wang2012,kumar2015,kumar2016,kumar2017}, jets emanating from the vicinity of the initial null point \citep{lynch2009,karpen2012,wyper2017,wyper2018,kumar2019}, plasmoids traveling along the BCS toward the footpoints \citep{karpen2012,guidoni2016}, bright EUV and soft X-ray emissions from heated plasma in the BCS and the side arcades \citep{kumar2013a}, stationary meterwave radio sources high in the corona \citep{aurass2013}, and pre-eruption microwave sources in the side arcades \citep{gary2018,karpen2019}. We emphasize that both timing and location are key factors in identifying such signatures as definitive evidence of the breakout mechanism, as we discuss below. 
  
\begin{figure*}
\centering{
\includegraphics[width=12cm]{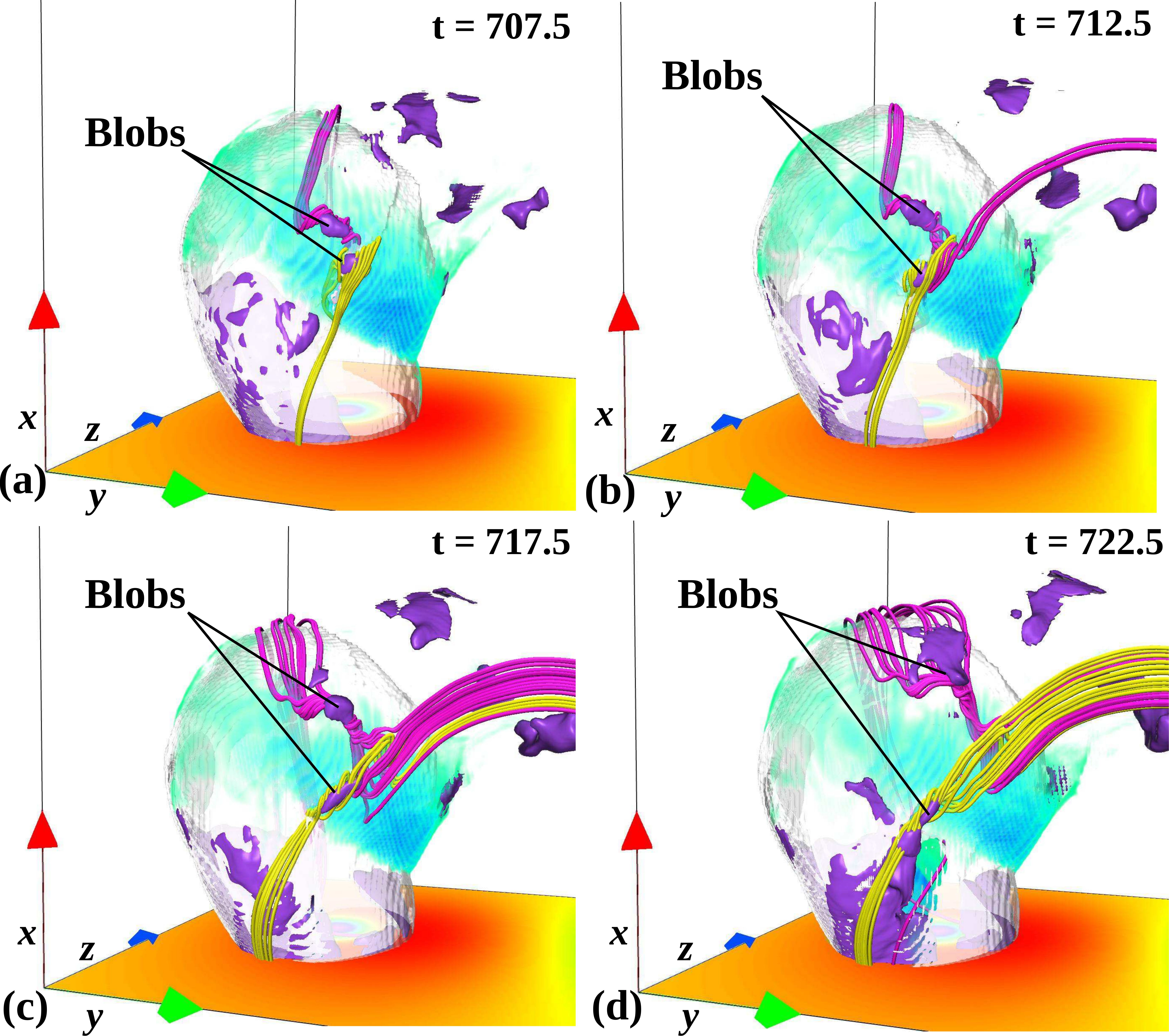}
}
\caption{Sequential snapshots of plasmoids --- enhanced-density blobs (purple isosurfaces) entrained in newly formed magnetic flux ropes (magenta and yellow lines) --- formed through breakout reconnection in a 3D MHD simulation of a coronal jet. The cyan volume shading shows the ratio of electric current to field strength (J/B) to highlight the breakout current layer. Adapted from Figure 15 of \citet{wyper2016b}, which includes an animation of the process of plasmoid formation and ejection.} 
\label{fig1}
\end{figure*}

Of these characteristic signs of breakout, all except the BCS plasmoids have been observed previously. Plasmoids are particularly significant because they offer a solution to the well-known fast-reconnection conundrum: plasmoids support a much higher reconnection rate for long current sheets \citep{loureiro2007,bhattacharjee2009,huang2013} than the much slower Sweet--Parker rate, which is too slow to match observations. Moreover, plasmoids can accelerate particles to X-ray and microwave-emitting energies \citep{drake2006,guidoni2016}. As shown in Figure \ref{fig1} and in the events discussed in this paper, the breakout current sheet can become extremely long before an eruption. In our simulations of resistive-kink and breakout jets, plasmoids always form in the BCS when reconnection is fast, facilitating the eruption \citep{karpen2012,wyper2016b,wyper2018}. High-resolution, high-cadence imaging is required to detect these small, dynamic features in the BCS. We report here the first direct detection of BCS plasmoids in two well-observed coronal jets, which meet the requirements and display the signatures of the resistive-kink and breakout models respectively. The detection of these plasmoids is observational proof that plasmoids are central features of fast reconnection in both two-dimensional (2D; flare) and three-dimensional (3D; null-point breakout) current sheets.

After describing the data selection, we present observations and analysis of two sequential jets from the same source region: the first without and the second with a filament eruption. We discuss our interpretation of these events and summarize the evidence for the breakout mechanism in both cases.

\begin{figure*}
\centering{
\includegraphics[width=5.8cm]{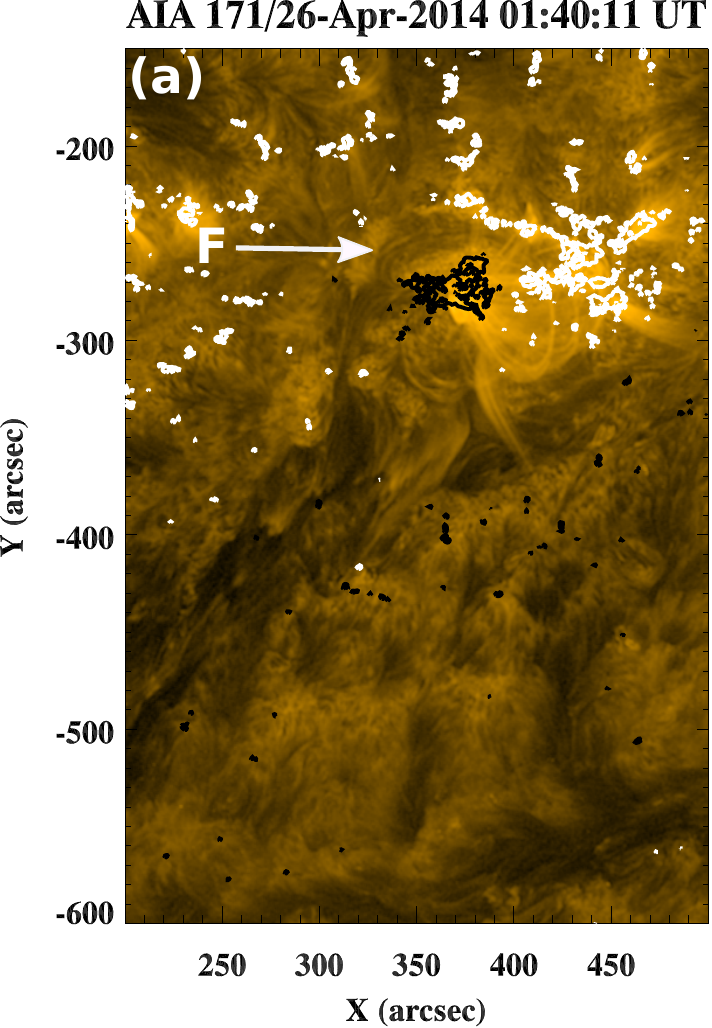}
\includegraphics[width=4.8cm]{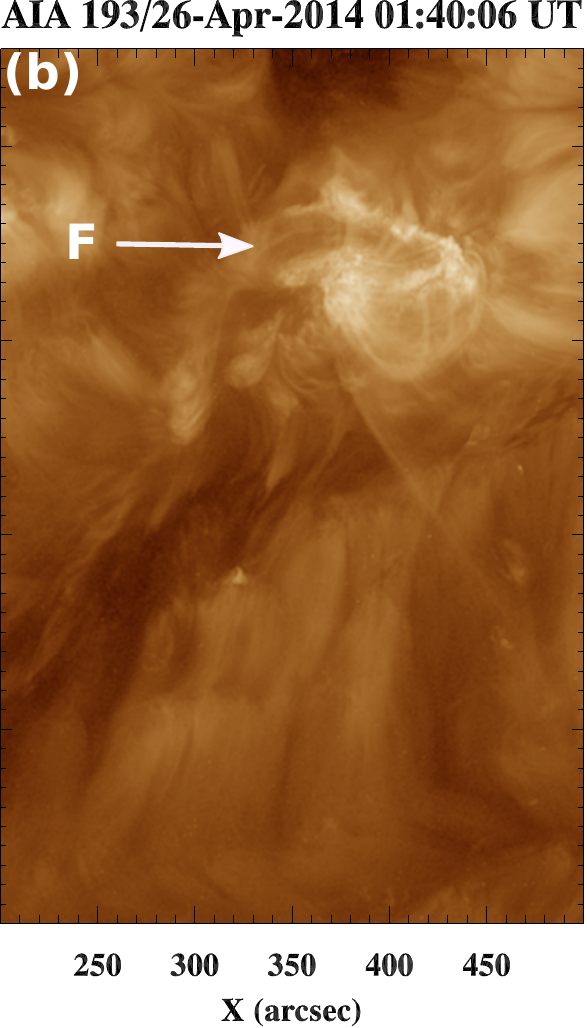}
\includegraphics[width=4.85cm]{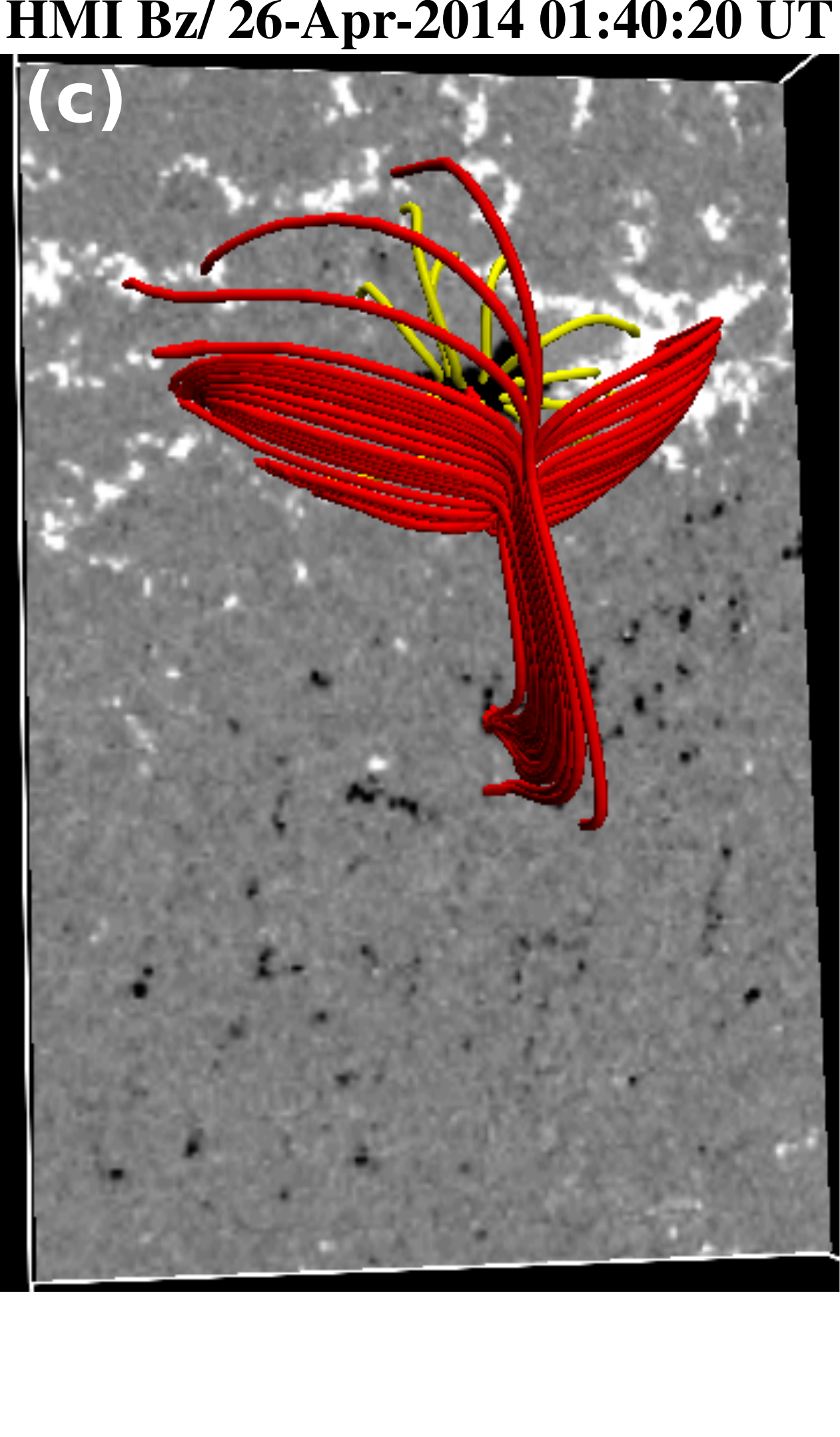}

\includegraphics[width=5.72cm]{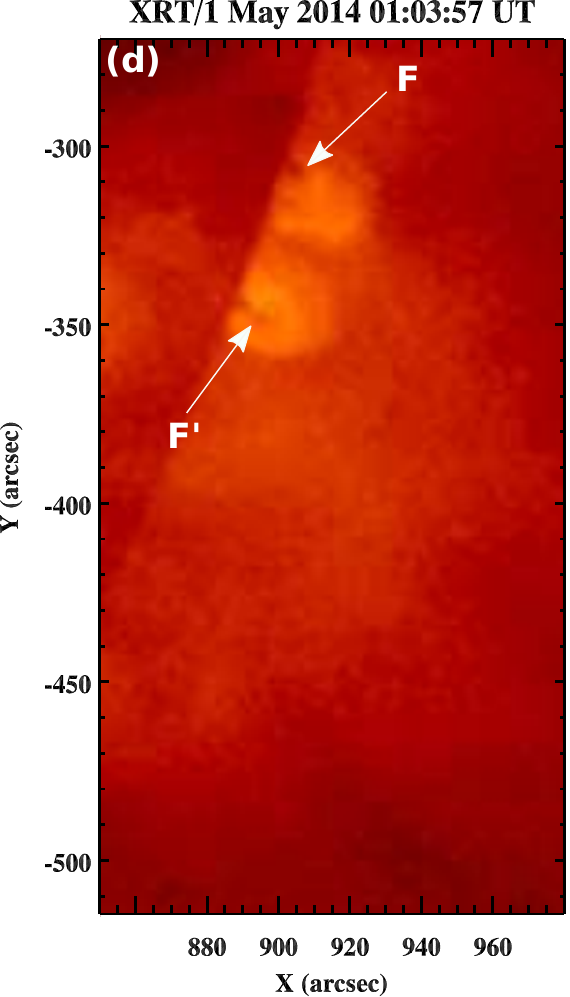}
\includegraphics[width=4.72cm]{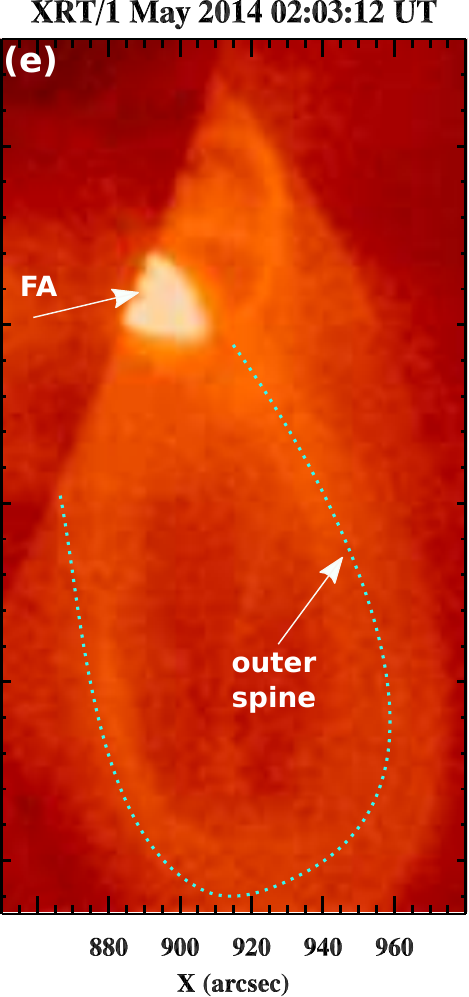}

}
\caption{Observations of AR 12044 four days before (a-c) and on the day of (d-e) the jets under discussion. (a,b) SDO/AIA 171 and 193~\AA\ images at 01:40:20 UT on 2014 April 26. The 171 \AA\ image is overlaid by the cotemporal HMI magnetogram contours of positive (white) and negative (black) polarities (levels=$\pm$200 G). (c) SDO/HMI magnetogram and selected field lines from a potential-field extrapolation in the jet source region at the same time as the EUV images in (a) and (b). Yellow field lines are closed beneath the fan, red field lines delineate the separatrix surface and closed spine. (d,e) Hinode/XRT images of the jet source region on 2014 May 1 shortly before the first jet (d) and after the second jet (e).  FA=flare arcade, and F and F$^\prime$ are filament segments on the same PIL within the AR.}
\label{fig2}
\end{figure*}

\begin{figure*}
\centering{
\includegraphics[width=18cm]{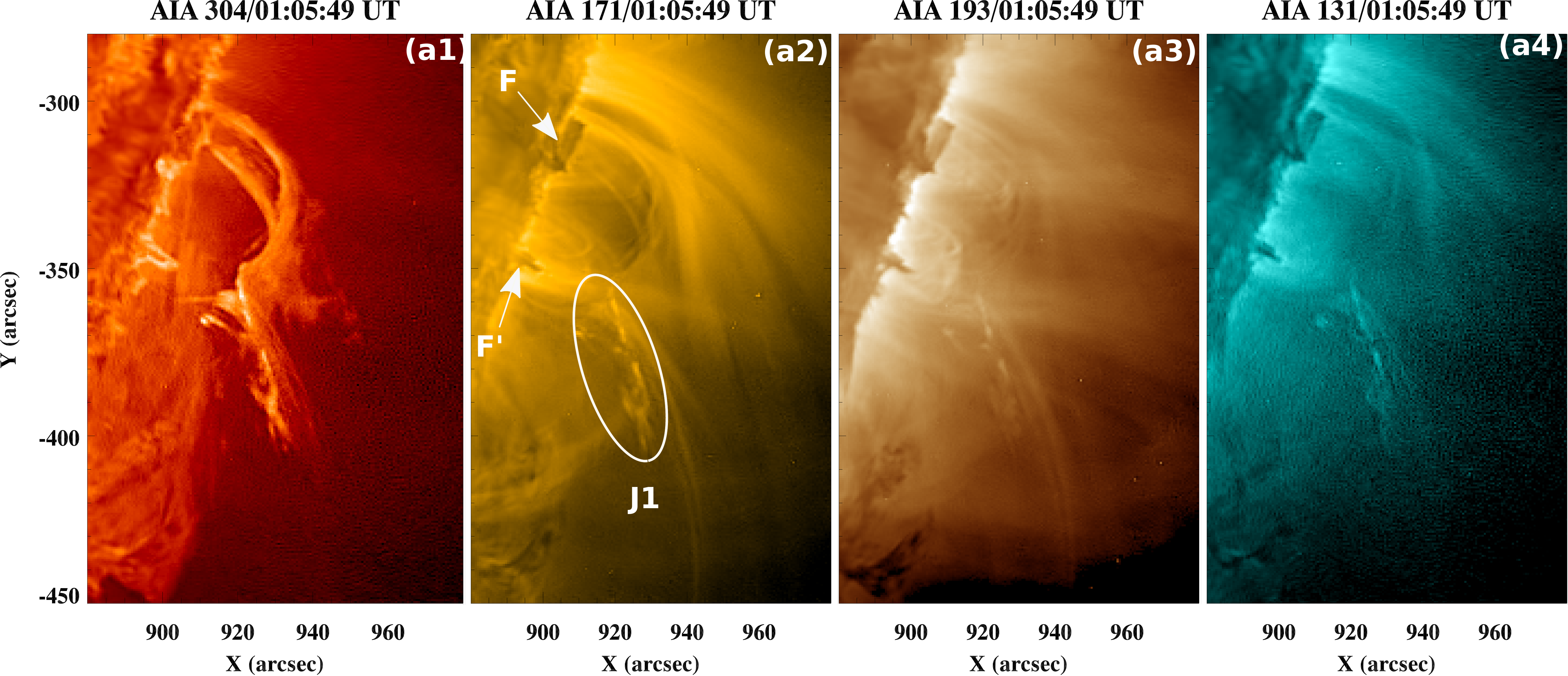}

\includegraphics[width=4.8cm]{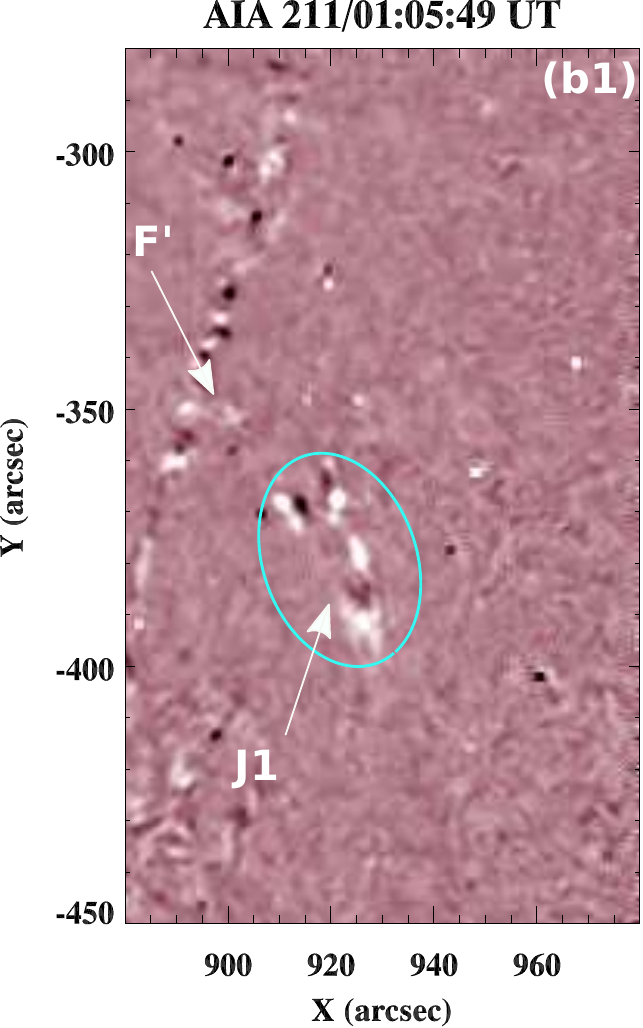}
\includegraphics[width=3.85cm]{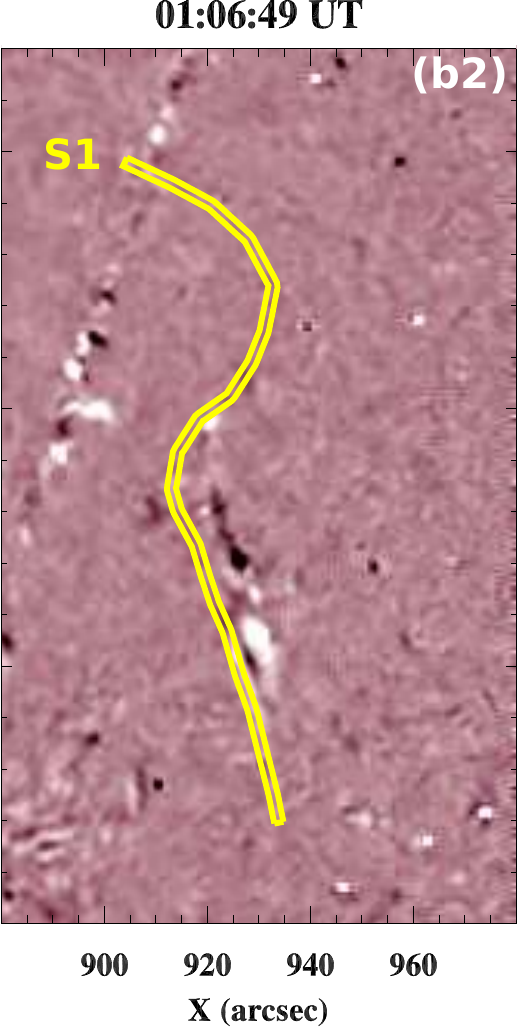}
\includegraphics[width=3.85cm]{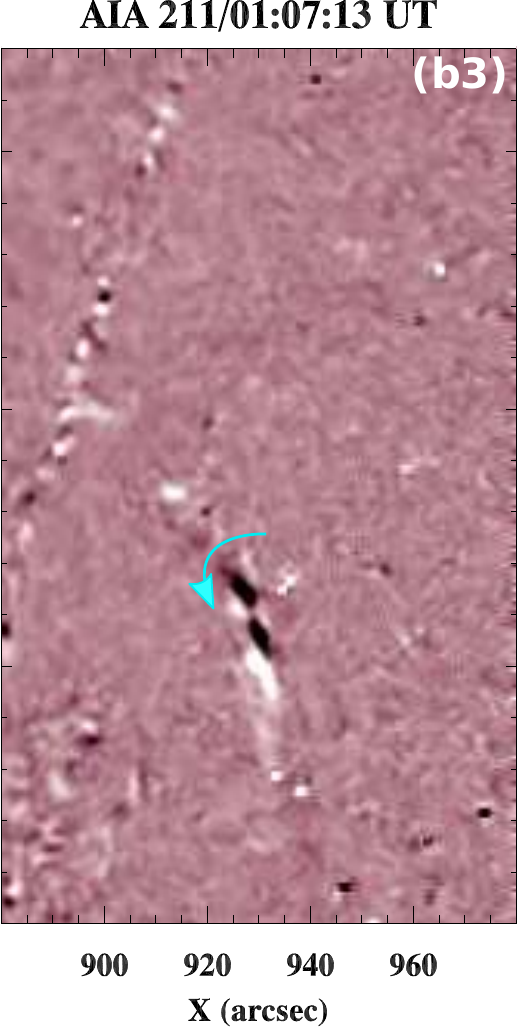}
\includegraphics[width=4.7cm]{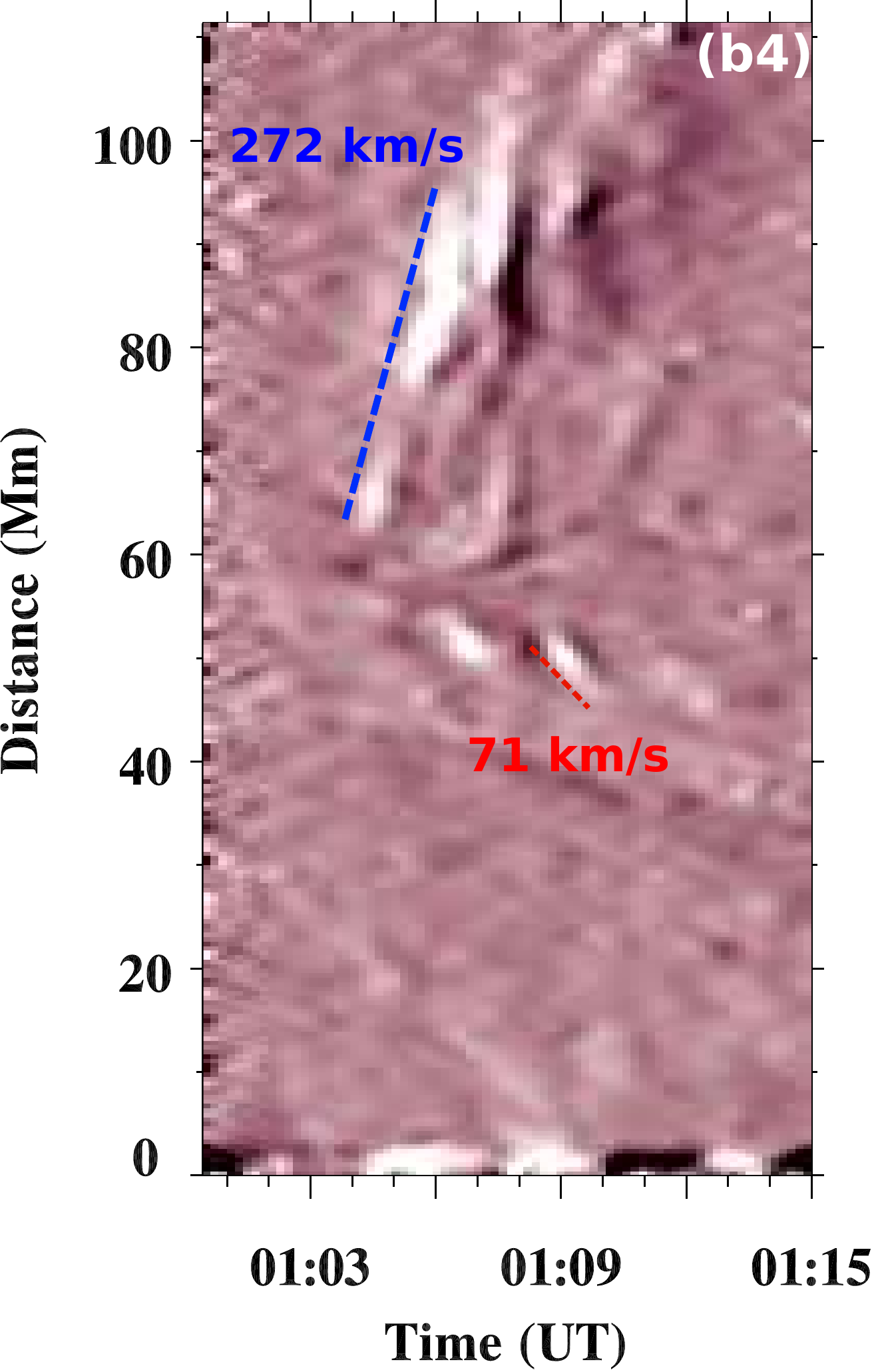}

\includegraphics[width=6.4cm]{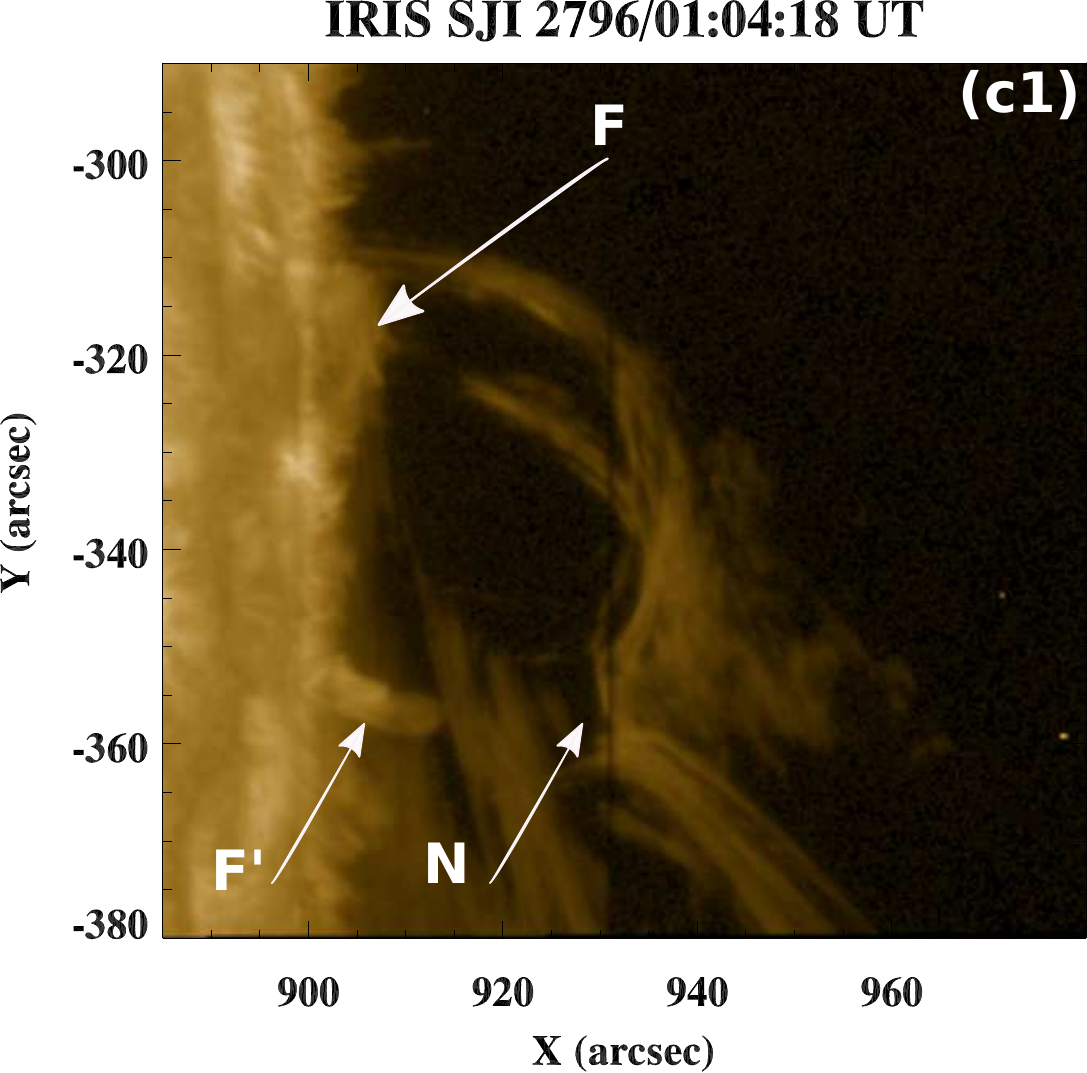}
\includegraphics[width=5.4cm]{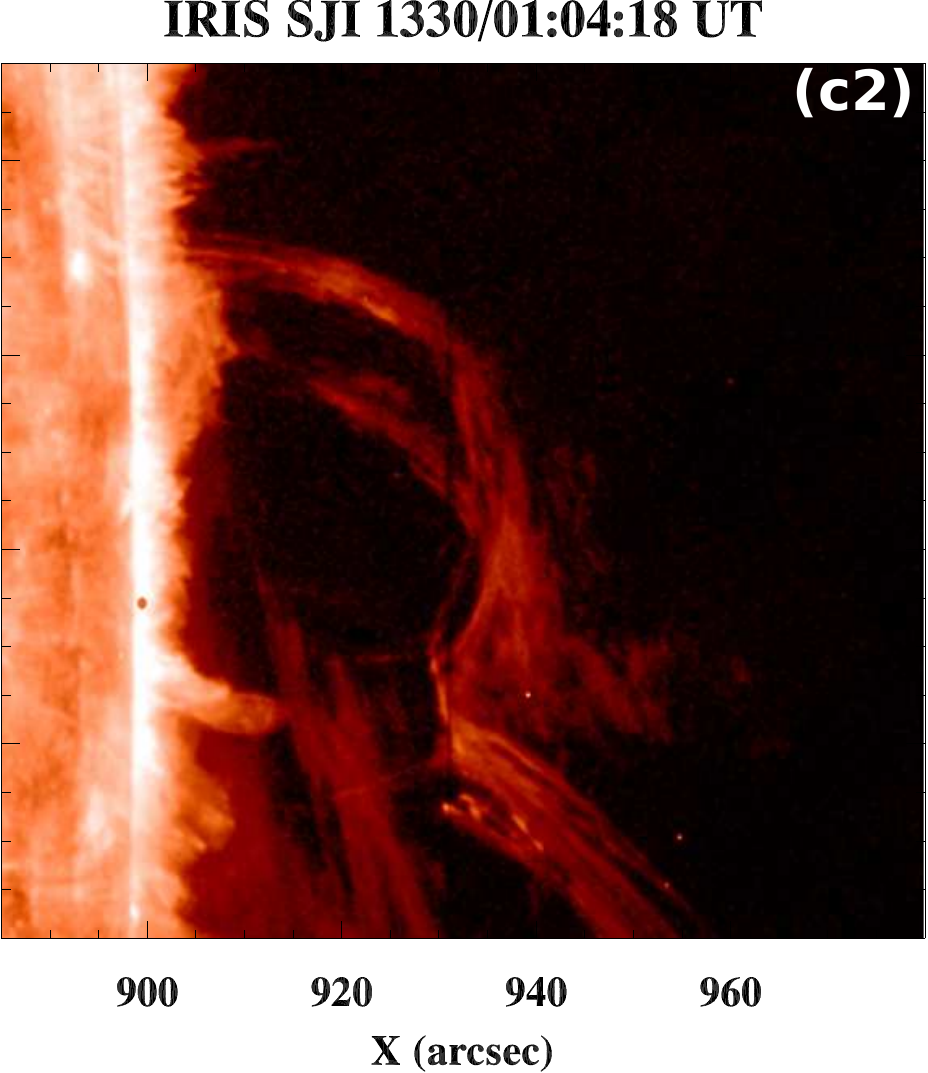}
\includegraphics[width=5.4cm]{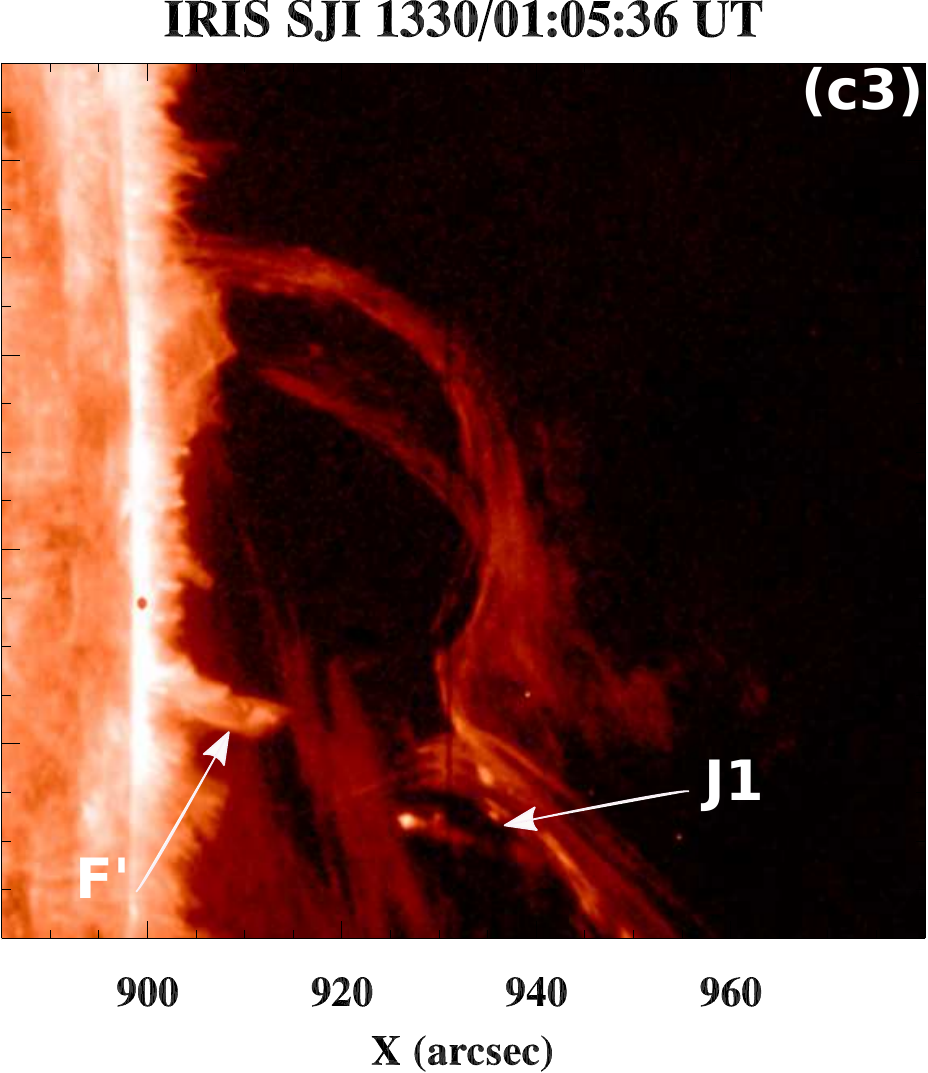}

}
\caption{SDO/AIA and IRIS observations of the first jet, J1. (a1-a4) SDO/AIA 304, 171, 193, and 131~\AA\ images at  01:05:49 UT; the white oval in (a2) surrounds J1. (b1-b3) AIA 171~\AA\ running-difference  ($\Delta$t=24 s) images at 3 times during J1 (cyan oval in (b1)). The yellow outlines show slit S1 used to create (b4), and the cyan arrow in (b3) marks the direction of the counterclockwise rotation of the jet. (b4) Time-distance plot of AIA 171~\AA\ intensity along slit S1 in (b2). (c1-c3) IRIS 2796~\AA\ (c1) and 1330~\AA\ (c2-c3) slit-jaw images before and during the onset of J1. N=null point, F and F$^\prime$ are segments of the same filament. (An animation of this figure covering both jets is available online).} 
\label{fig3}
\end{figure*}

%
\begin{figure*}
\centering{
\includegraphics[width=4.1cm]{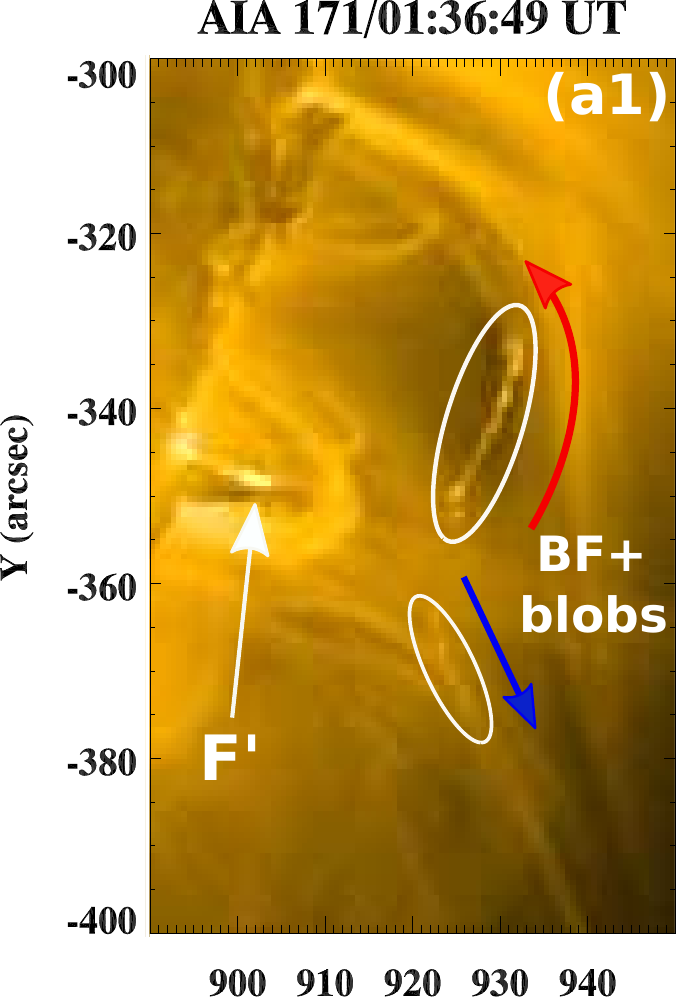}
\includegraphics[width=3.2cm]{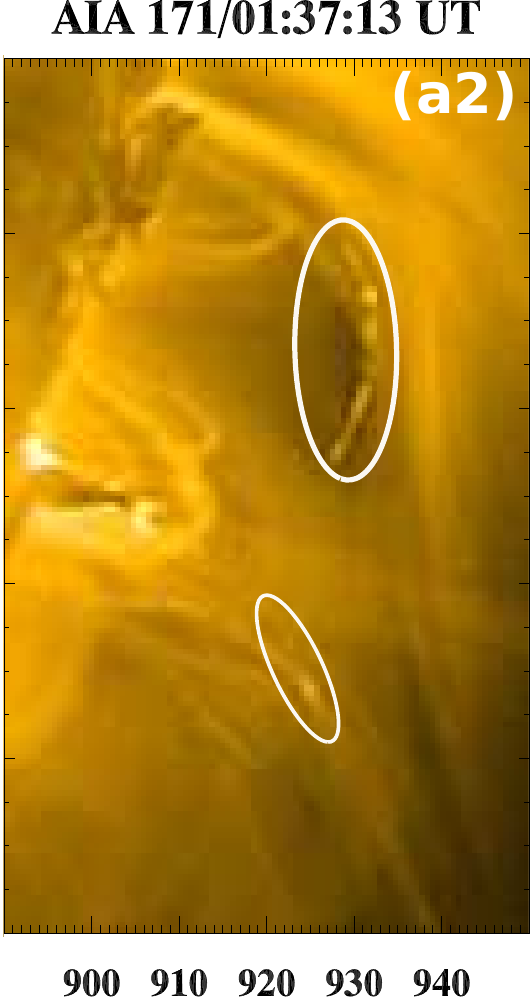}
\includegraphics[width=3.2cm]{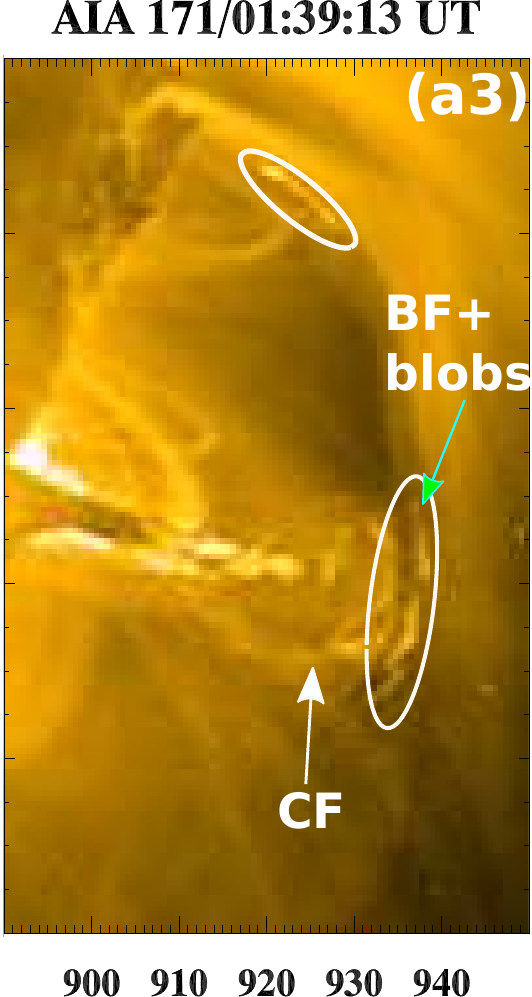}
\includegraphics[width=3.2cm]{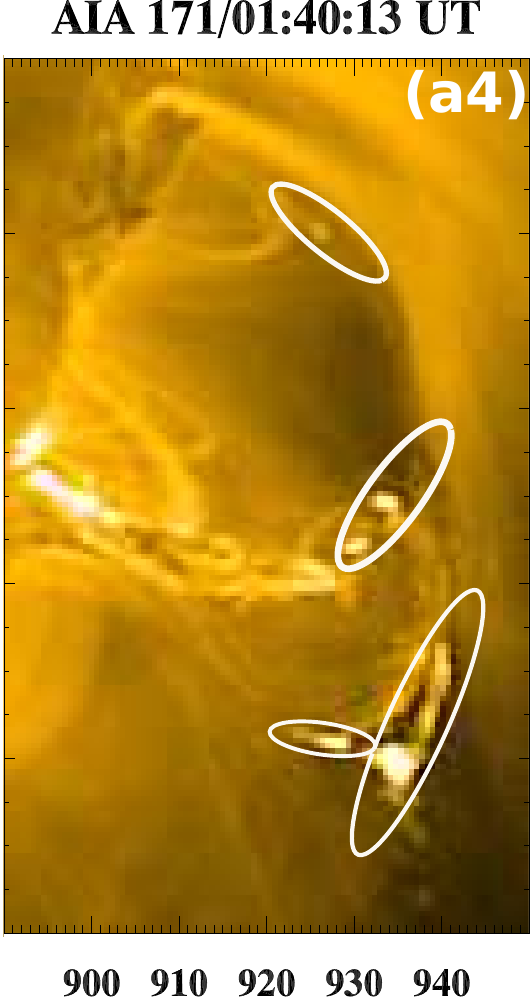}
\includegraphics[width=3.2cm]{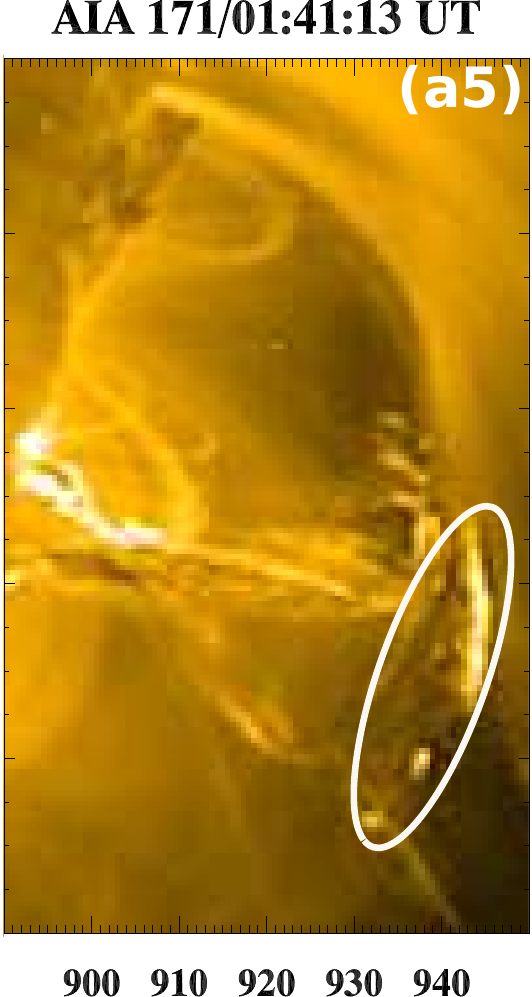}
\vspace{0.3cm}

\includegraphics[width=4.1cm]{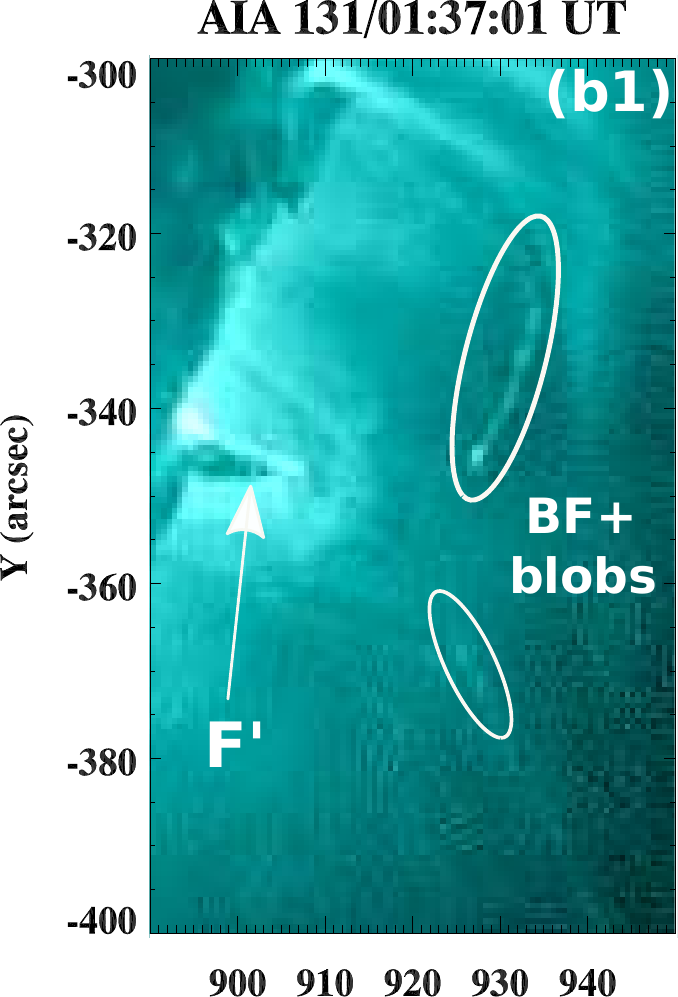}
\includegraphics[width=3.2cm]{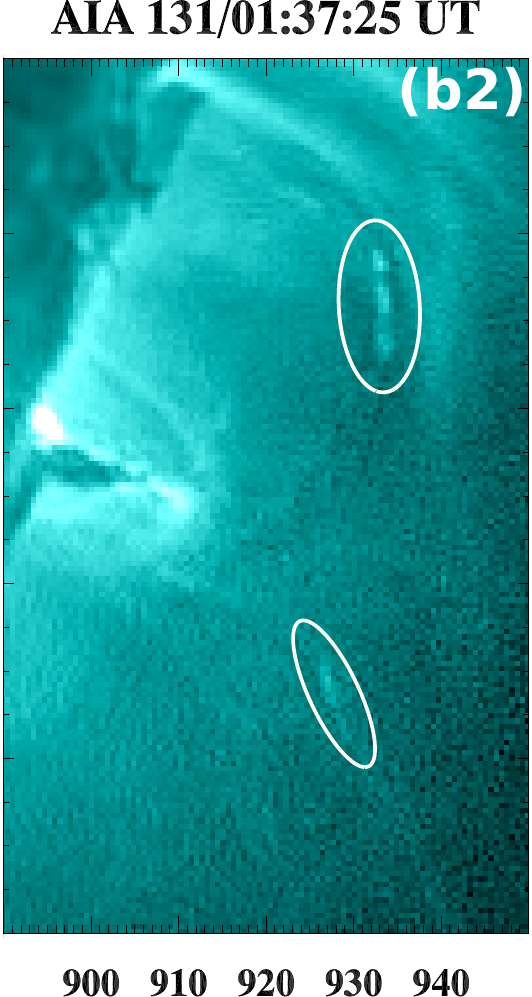}
\includegraphics[width=3.2cm]{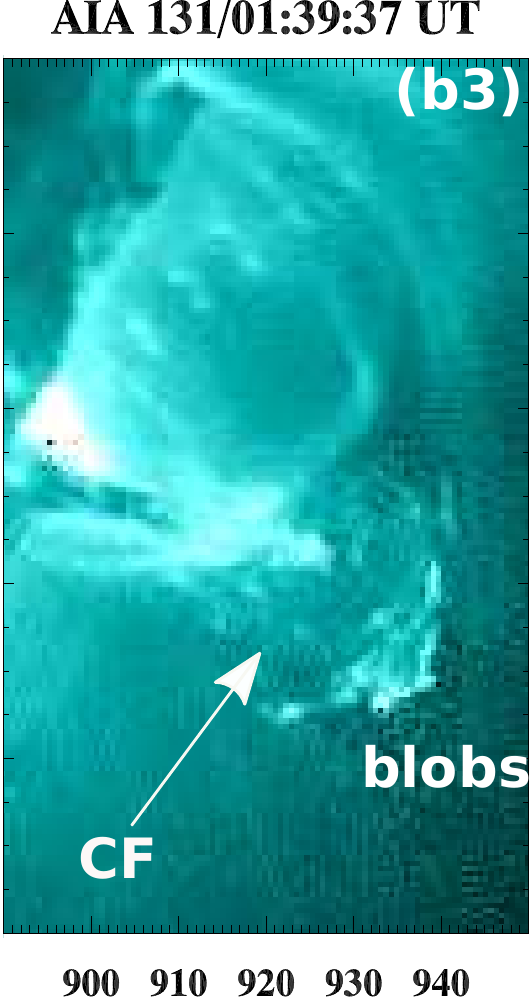}
\includegraphics[width=3.2cm]{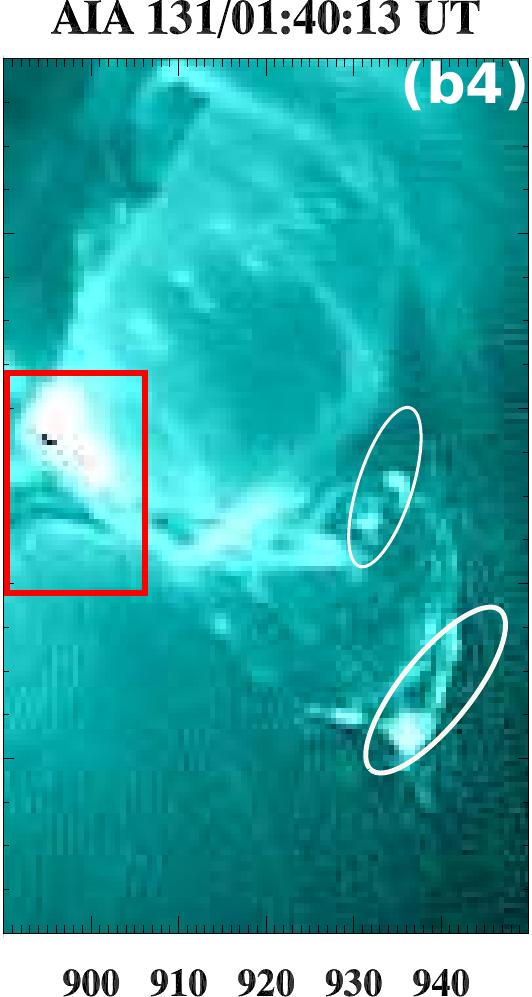}
\includegraphics[width=3.2cm]{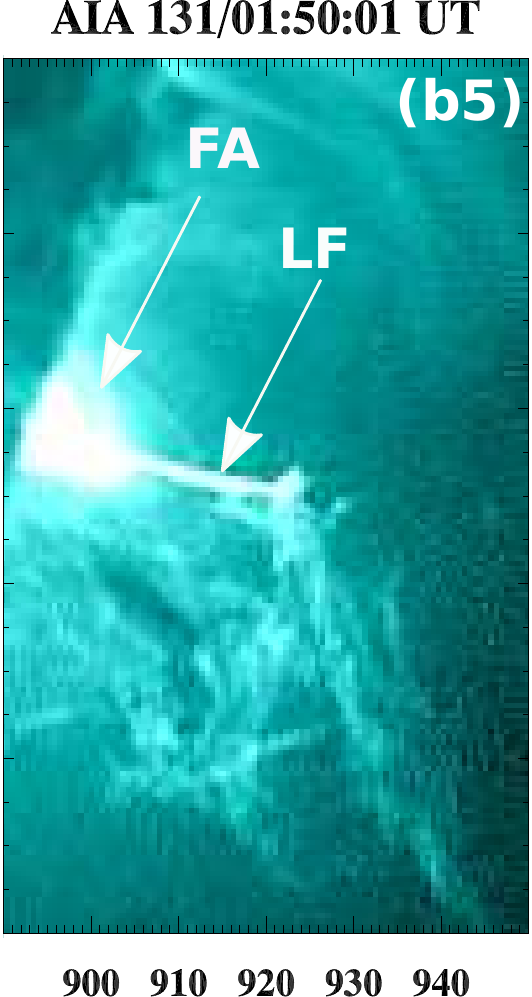}
\vspace{0.3cm}

\includegraphics[width=4.1cm]{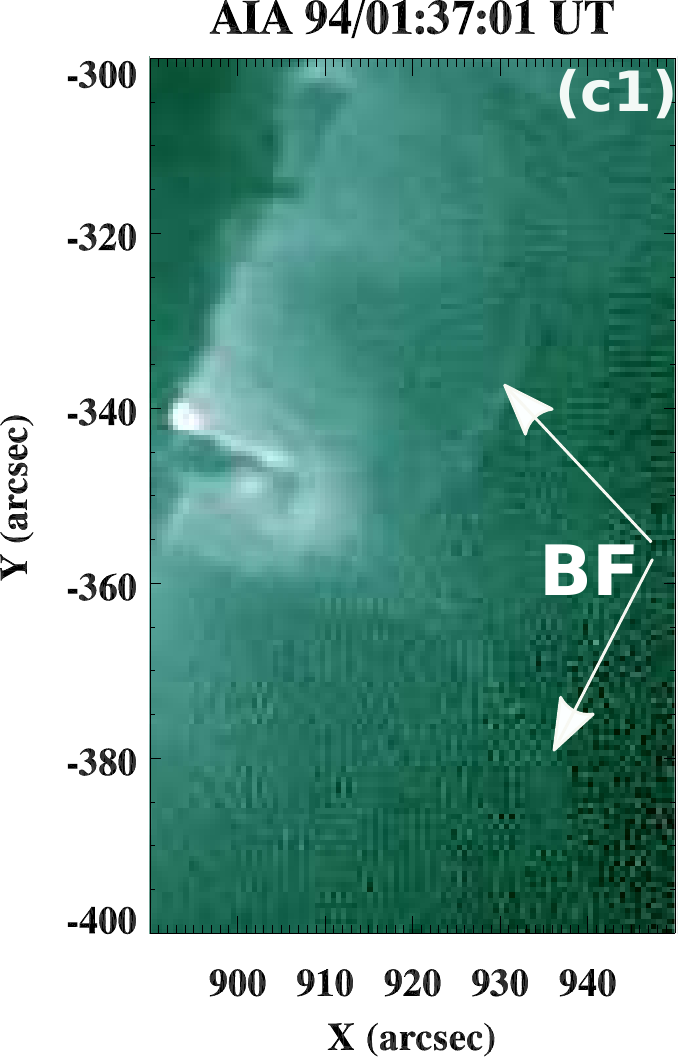}
\includegraphics[width=3.2cm]{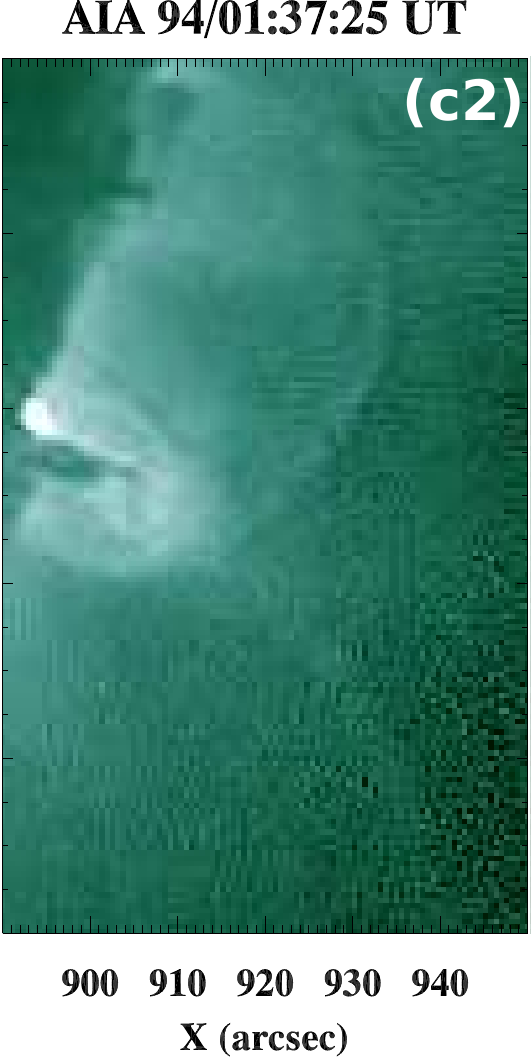}
\includegraphics[width=3.2cm]{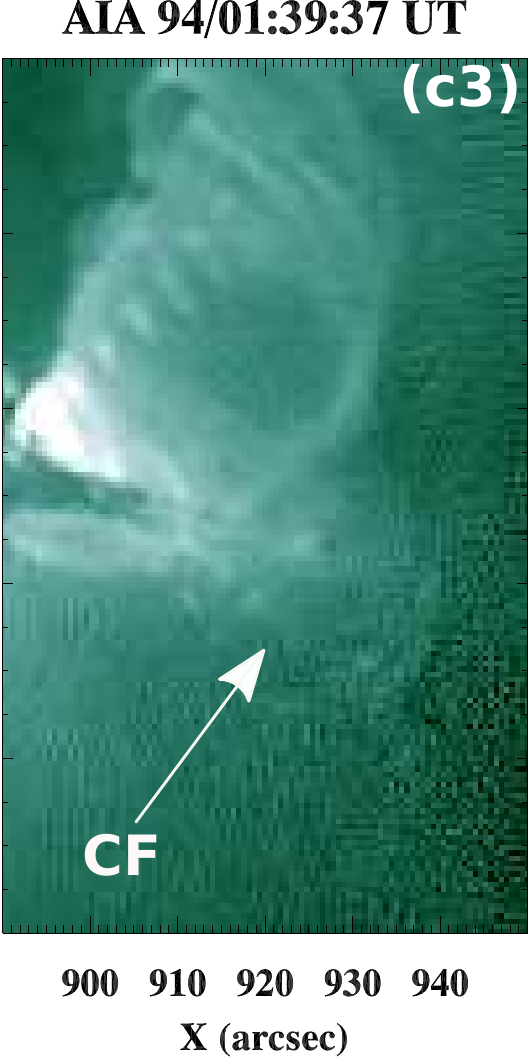}
\includegraphics[width=3.2cm]{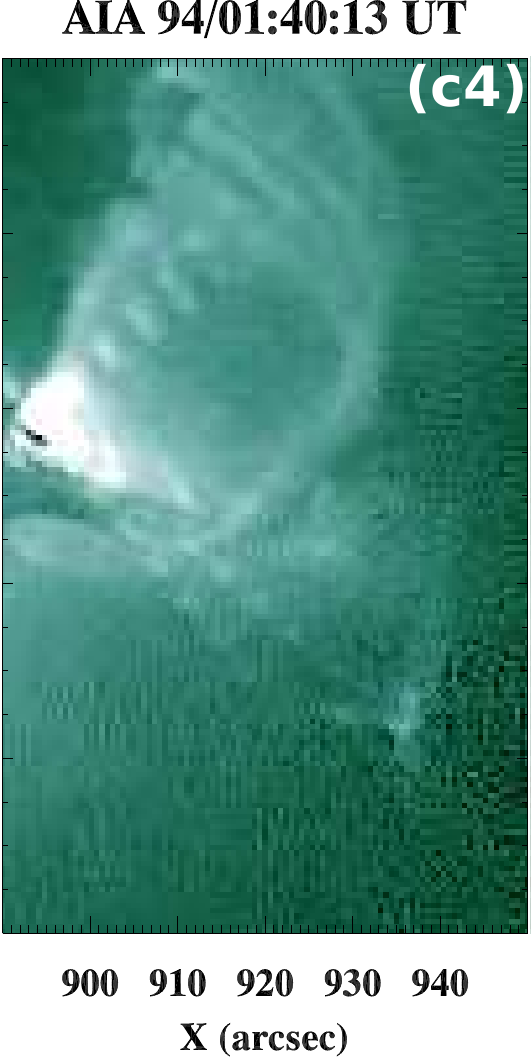}
\includegraphics[width=3.2cm]{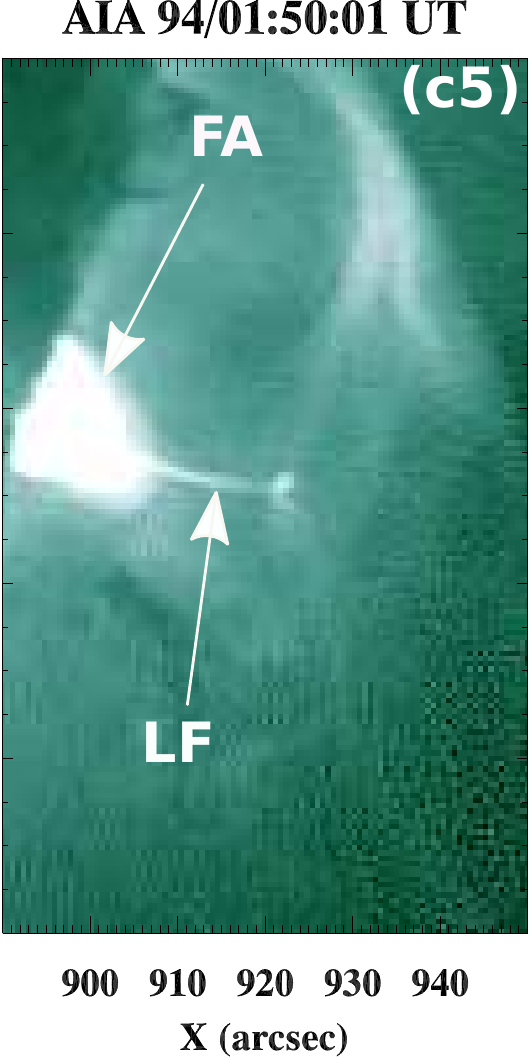}
}
\caption{SDO/AIA observations of the second jet, J2. (a1-a5) SDO/AIA 171~\AA, (b1-b5) 131~\AA, and (c1-c5) 94~\AA\ images. (BF=bidirectional flows, CF=circular feature, LF=linear feature). Other notations are defined in previous figures. (An animation of the AIA 131 and 94~\AA\ channels is available online; an animation of the AIA 171~\AA\ channel is included in the animation accompanying Figure \ref{fig3}.)} 
\label{fig4}
\end{figure*}


\section{Observations}\label{obs} 
We analyzed {\it Solar Dynamics Observatory} (SDO)/Atmospheric Image Assembly (AIA; \citealt{lemen2012}) full-disk images of the Sun (field-of-view $\sim$1.3~R$_\odot$) with a spatial resolution of 1.5$\arcsec$ (0.6$\arcsec$~pixel$^{-1}$) and a cadence of 12~s, in the following channels: 304~\AA\ (\ion{He}{2}, at temperature $T\approx 0.05$~MK), 171~\AA\ (\ion{Fe}{9}, $T\approx 0.7$~MK), 193~\AA\ (\ion{Fe}{12}, \ion{Fe}{24}, $T\approx  1.2$~MK and $\approx 20$~MK), 211~\AA\ (\ion{Fe}{14}, $T\approx 2$~MK), AIA 94~\AA\ (\ion{Fe}{10}, \ion{Fe}{18}, $T\approx$1 MK, $T\approx$6.3 MK), and 131~\AA\ (\ion{Fe}{8}, \ion{Fe}{21}, \ion{Fe}{23}, i.e., 0.4, 10, 16 MK) images. A 3D noise-gating technique \citep{deforest2017} was used to clean the AIA images. AIA 211 \AA\ running-difference images with $\Delta t = 24$ s (Figures \ref{fig3}(b1-b3), \ref{fig5}(a-f), and accompanying movie) and time-distance plots along selected slots imposed on the same running-difference images (Figure \ref{fig3}(b4) and Figure \ref{fig5}(g-i)) revealed faint features in the dynamically evolving magnetic structure that are difficult to discern in the undifferenced images, and enabled the measurement of the projected speeds of key dynamic features (note that the actual speeds are likely to be higher). 

Selected images from the {\it Hinode} X-Ray Telescope (XRT; \citealt{golub2007}) were used to understand the magnetic topology of the source region. We inspected magnetograms from the SDO/Helioseismic and Magnetic Imager (HMI; \citealt{schou2012}) for the magnetic context, and found little evolution during the period of interest.  We also utilized Interface Region Imaging Spectrograph (IRIS) slit-jaw images (19-s cadence, 0.167$\arcsec$~pixel$^{-1}$) of the jet source region in 1330~\AA\ (\ion{C}{2}, \ion{Fe}{21}, log T=3.7-7.0 K) and 2796~\AA\ (\ion{Mg}{2} h/k, log T=3.7-4.2 K) channels \citep{de-pontieu2014}. The jets and associated structures identified in the AIA data were not always clear in the IRIS 2796 \AA\ channel, but some were detected in the hotter 1330 \AA\ channel.

\subsection{Pre-event Environment}\label{bkg}

The jets occurred in active region (AR) NOAA 12044 (S21W38) on 2014 May 1, when the AR was located near the western limb. This plage region lacked sunspots and was situated at the outer boundary of a small, positive-polarity equatorial coronal hole. A few days earlier, the AR was located near disk center and thus was well observed by both the SDO/AIA and HMI. As shown in Figure \ref{fig2}(a)-(c), a curved filament (F) lies along the northern section of the polarity inversion line (PIL), where the adjacent flux of both polarities was stronger than around the southern half of the PIL. A few diffuse loops connected the central minority-polarity region to the remote opposite polarity on the southern side of the AR. As the AR rotated toward the limb, a filament (F$^{\prime}$) formed along the southern portion of the same PIL. Although these filaments reside in the same filament channel, we refer to them individually because they are visible at separate latitudes on the west limb (Figure \ref{fig2}(d)). In addition, their eruption history differs substantially, as the northern filament F was unaffected by either jet. 

A potential-field extrapolation from an HMI magnetogram on 2014 April 26, about five days before the eruptions, reveals a classic fan-spine topology: an asymmetric ``anemone'' configuration with bright loops (red) connecting mainly to the south and a closed outer spine also directed southward (Figure \ref{fig2}(c)). The yellow field lines under the red fan surface in the Figure are closed field lines above the filament.  AIA 171 and 193 \AA\ images on the same day show fan loops connecting from the central minority polarity (negative) to the surrounding majority polarity (positive) regions (Figure \ref{fig2}(a,b)), consistent with the extrapolated fan-spine structure. The Hinode X-ray Telescope (XRT) images of the AR on 2014 May 1 shortly before the first jet (at 01:03:57 UT) and after the second jet (at  02:03:12 UT) also reveal a classic embedded-bipole structure: a 3D null, closed loops beneath the fan surface, and a closed outer spine connecting to the south (Figure \ref{fig2}(d,e)). 

Interface Region Imaging Spectrograph (IRIS) slit-jaw images in 2796 and 1330 \AA\ (Figure \ref{fig3}(c1-c3) and accompanying movie) show copious, persistent coronal rain originating near the null about 3 hours before the first jet. From 10 UT on April 30 ($\approx$15 hours before the first jet) until the explosive second jet, weaker rain fell along the northern fan surface. The classic fan-spine topology is distinctly outlined by the coronal rain because the cool plasma follows the fan surface (see accompanying IRIS movie). A possible explanation for the creation of persistent coronal rain in this magnetic topology is interchange reconnection at the null \citep{mason2019}, which is significant in the context of our interpretation of the jets.  
Given the strong evidence for the fan-spine configuration underlying this eruptive event, we will assume this fundamental geometry in our presentation of the observations and particularly in our subsequent interpretation. 

\subsection{Jet 1}\label{j1}

The first jet (J1) began at $\approx$01:04:01 UT and stopped ejecting new material around 01:10 UT (see AIA movie). J1 originated near the null point at the intersection of fan and spine, and propagated along the outer spine toward a remote site south of the AR (Figure \ref{fig2}(e,f); also see AIA 94 \AA\ channel base and running difference movie). Multiple tiny plasma blobs, each $\ge$2$\arcsec$ across, were detected in this jet at about 01:05:25-01:05:37 UT, while similarly sized plasma blobs propagated from the null along the northern fan loops (Figure \ref{fig3}(a1-a4)). We determined the counterclockwise direction of rotation of the outflow (cyan arrow in Figure \ref{fig3}(b3)) by tracking the dark features of J1 in the AIA 211 ~\AA\ running difference movie.  The jet speed, estimated from the time-distance plot extracted from slit S1 (Figure \ref{fig3}(b4)), was $\approx$272$\pm$16 \kms; the associated downflow speed of the fan blobs along the same slit was $\approx$71$\pm$8 \kms. 

Very little chromospheric activity, and no recognizable flare arcade, appeared in any of the AIA and IRIS bandpasses during J1 (Figure \ref{fig3}). A bright rim bordered the filament F$^{\prime}$ from $\approx$01:04 to 01:12 UT in all AIA bandpasses except 131~\AA, and bright points appeared in 304 \AA\ along the adjacent limb. Part of F$^{\prime}$ rose slightly (about 2-3 Mm) and then descended to its original height between 01:00 and 01:08 UT. Because the filament is too small and low to be adequately resolved by AIA, this activation is only visible in the IRIS observations (Figure \ref{fig3}(c1-c3) and accompanying IRIS movie). 

\begin{figure*}
\centering{
\includegraphics[width=4.85cm]{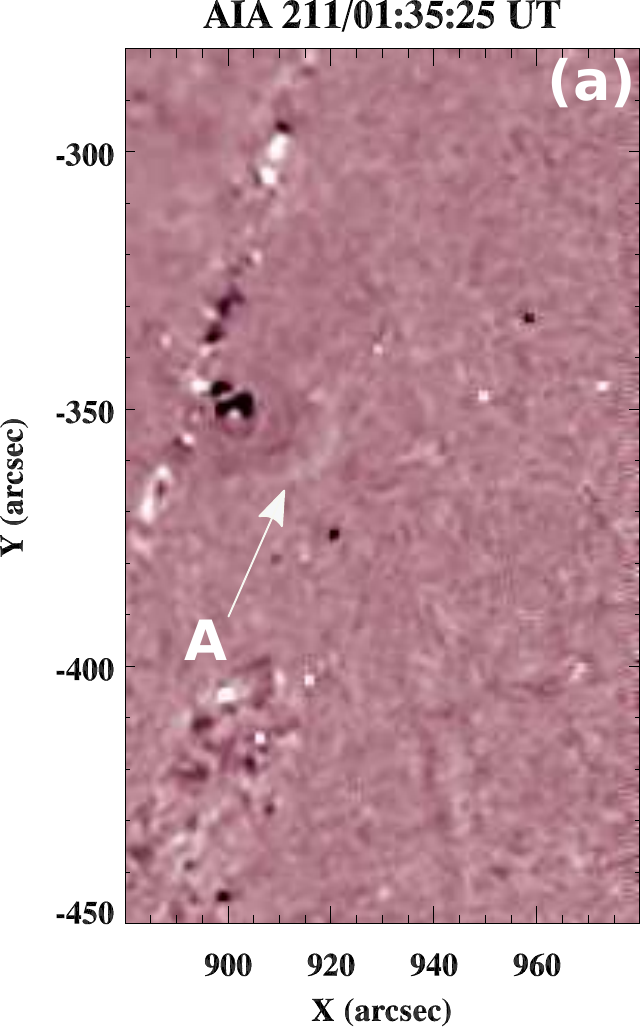}
\includegraphics[width=3.9cm]{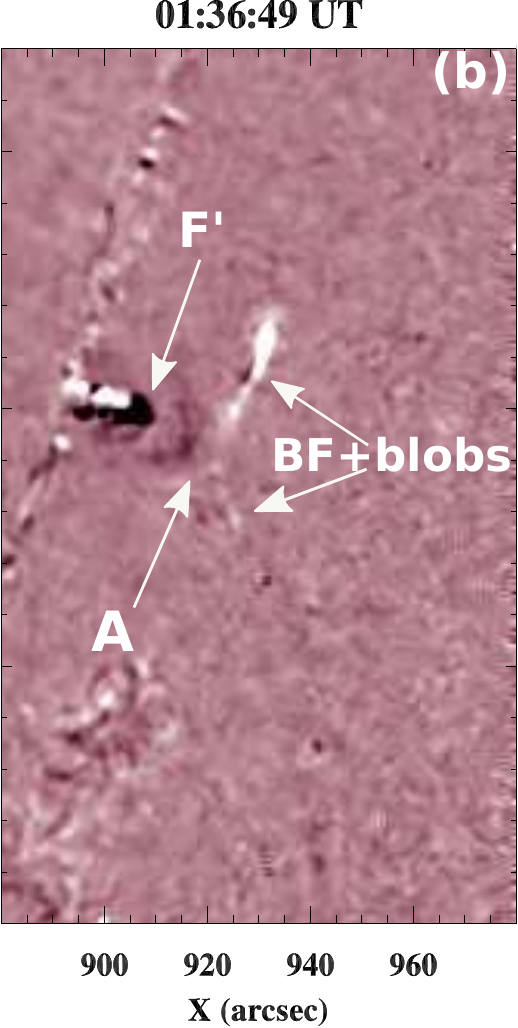}
\includegraphics[width=3.9cm]{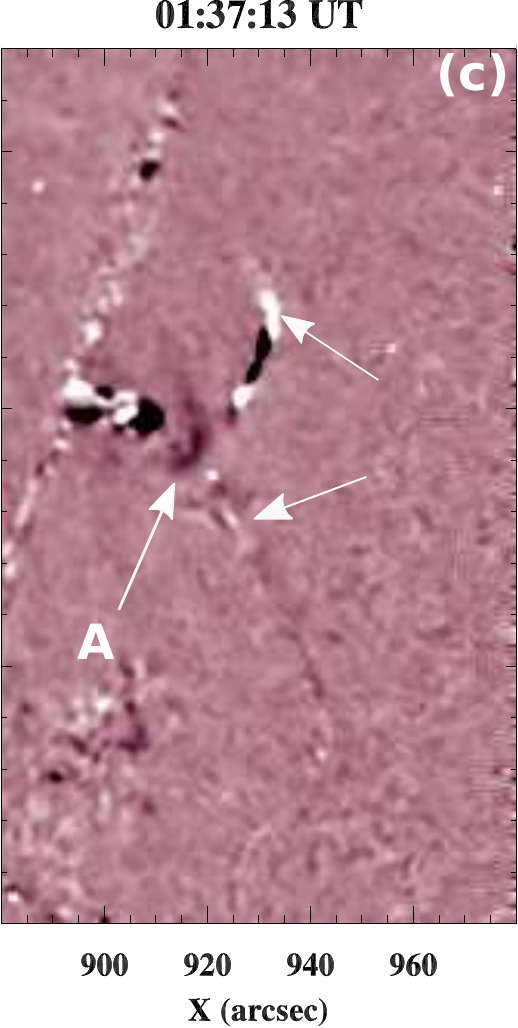}
\includegraphics[width=3.9cm]{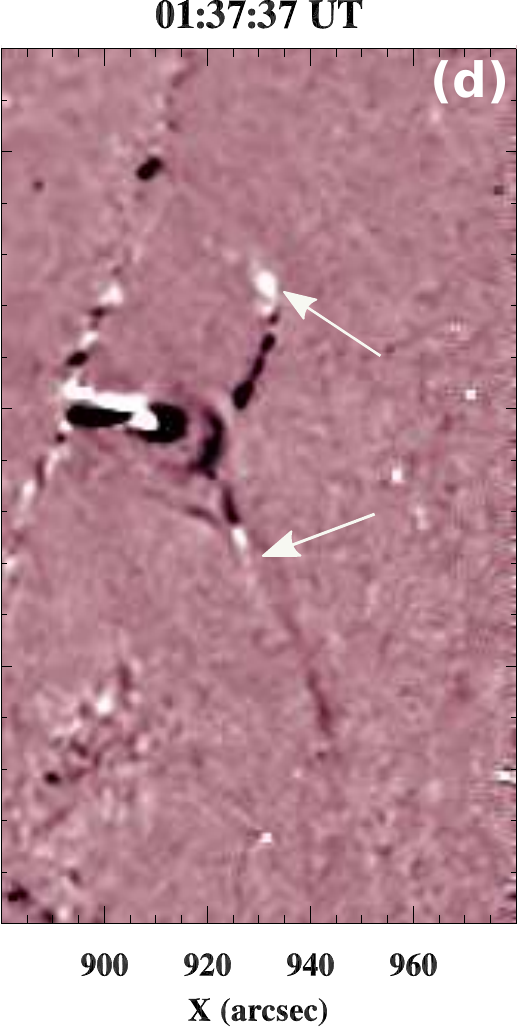}
\vspace{0.3cm}

\includegraphics[width=4.96cm]{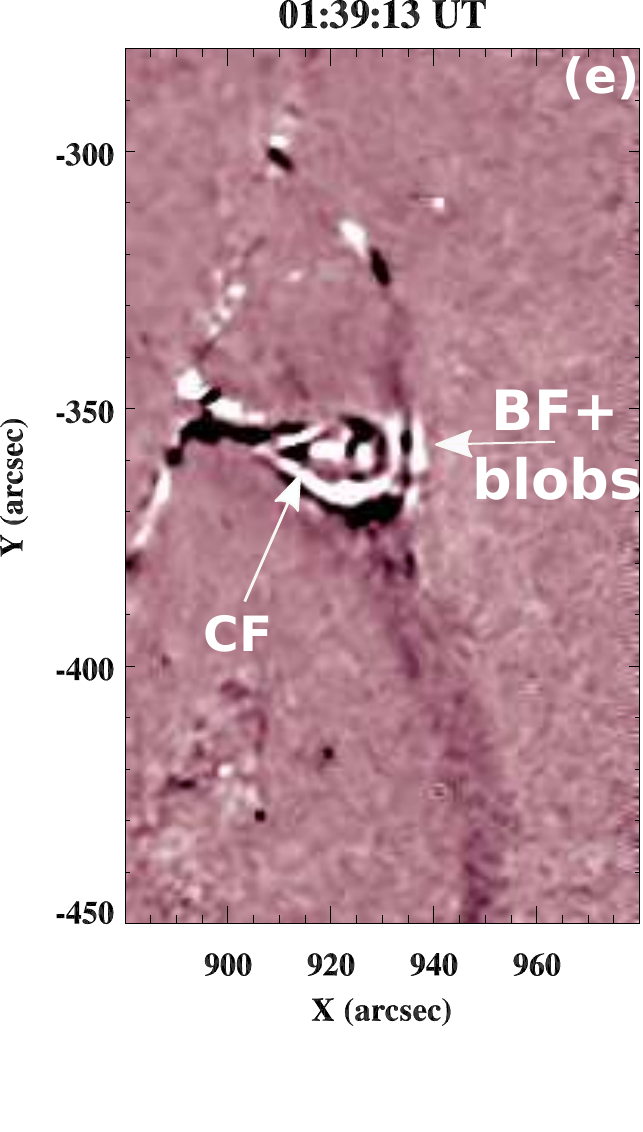}
\includegraphics[width=4.0cm]{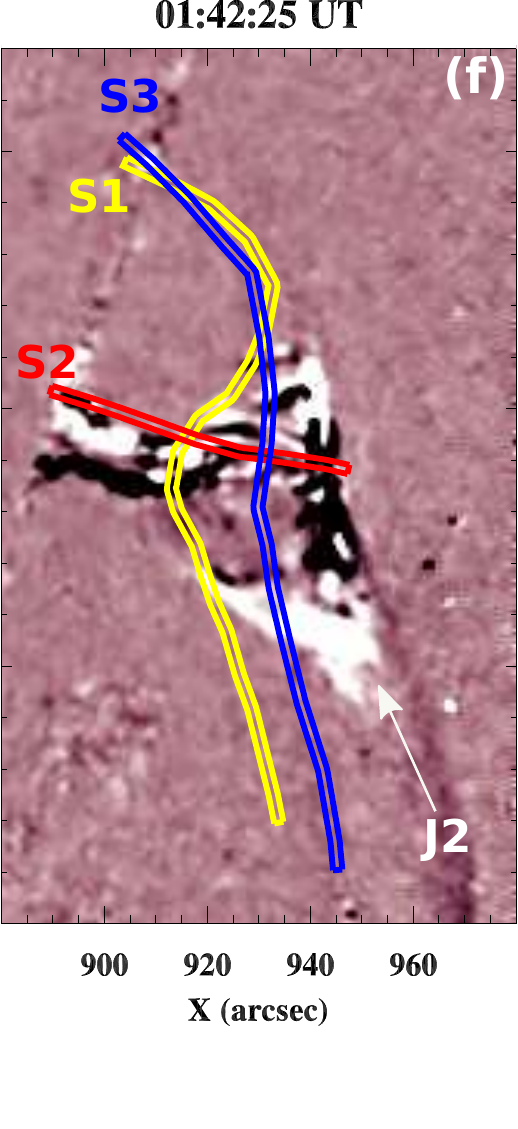}
\includegraphics[width=8.0cm]{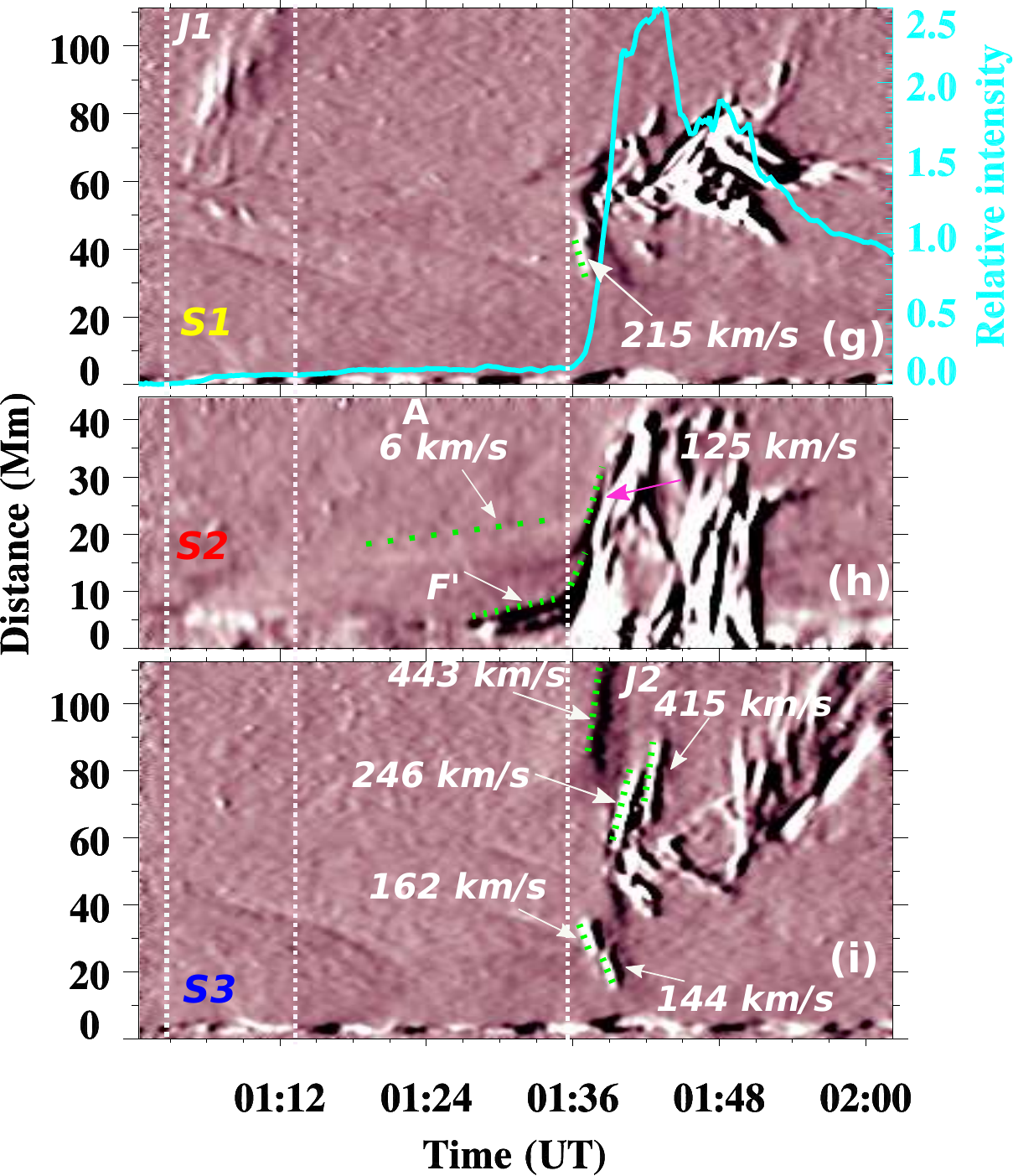}
}
\caption{A sequence of AIA 211~\AA\ images of the second jet, J2, showing the breakout current sheet, blobs, and bidirectional flows. (a-f) Selected 211~\AA\ running difference ($\Delta$t=24 s) images. (g-i) Time-distance intensity maps extracted from the 211 \AA\ running difference images along slices S1, S2, and S3, respectively, marked in panel (f). The cyan curve in (g) is the average intensity extracted from a box around the flare arcade using AIA 131~\AA\ base ratio images. The first two vertical dotted lines indicate the duration of J1, while the third vertical dotted line shows the onset time of J2. The speeds of distinct moving features, projected into the sky plane, are marked. Notations are defined in previous figures, except for arc A in (a) (see text for details). (An animation of this figure is available online as part of the animation accompanying Figure \ref{fig3}).} 
\label{fig5}
\end{figure*}

\begin{figure*}
\centering{

\includegraphics[width=6.4cm]{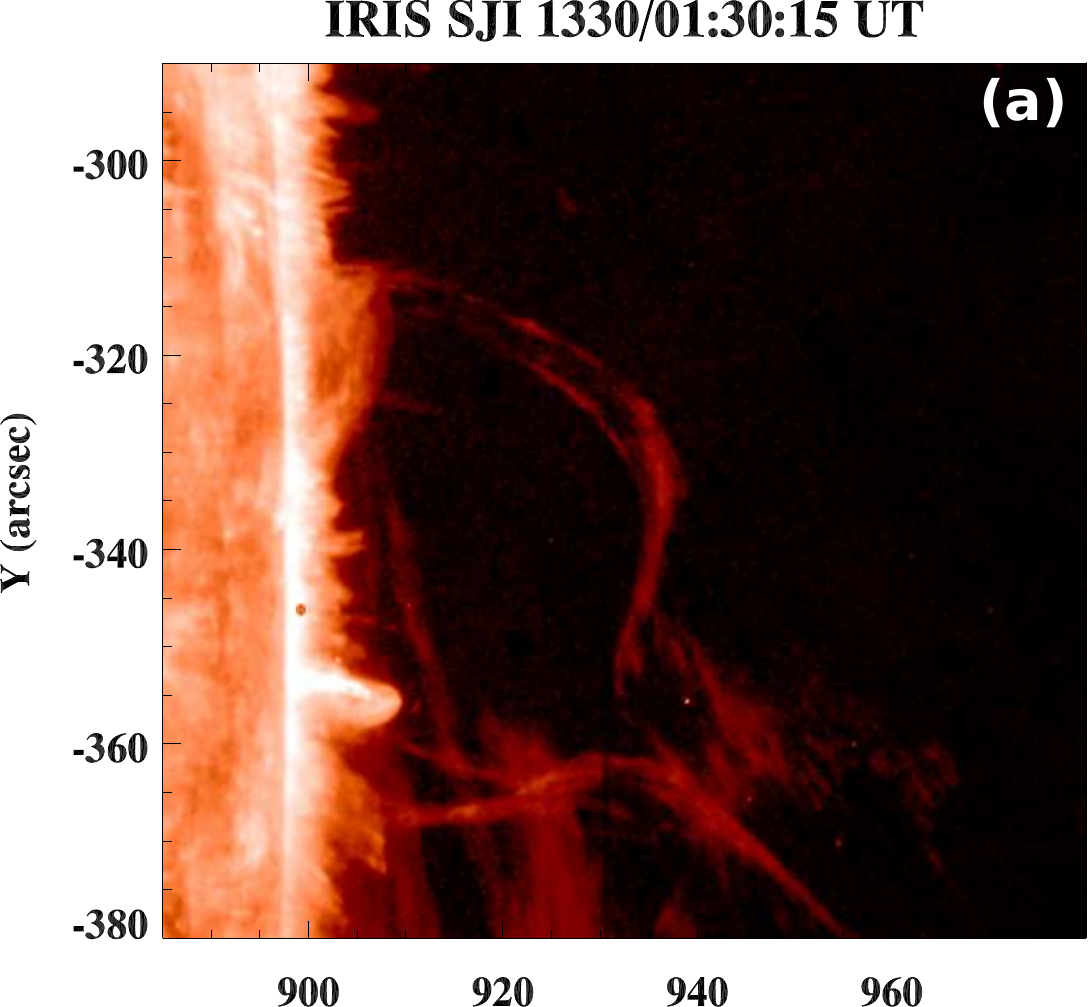}
\includegraphics[width=5.4cm]{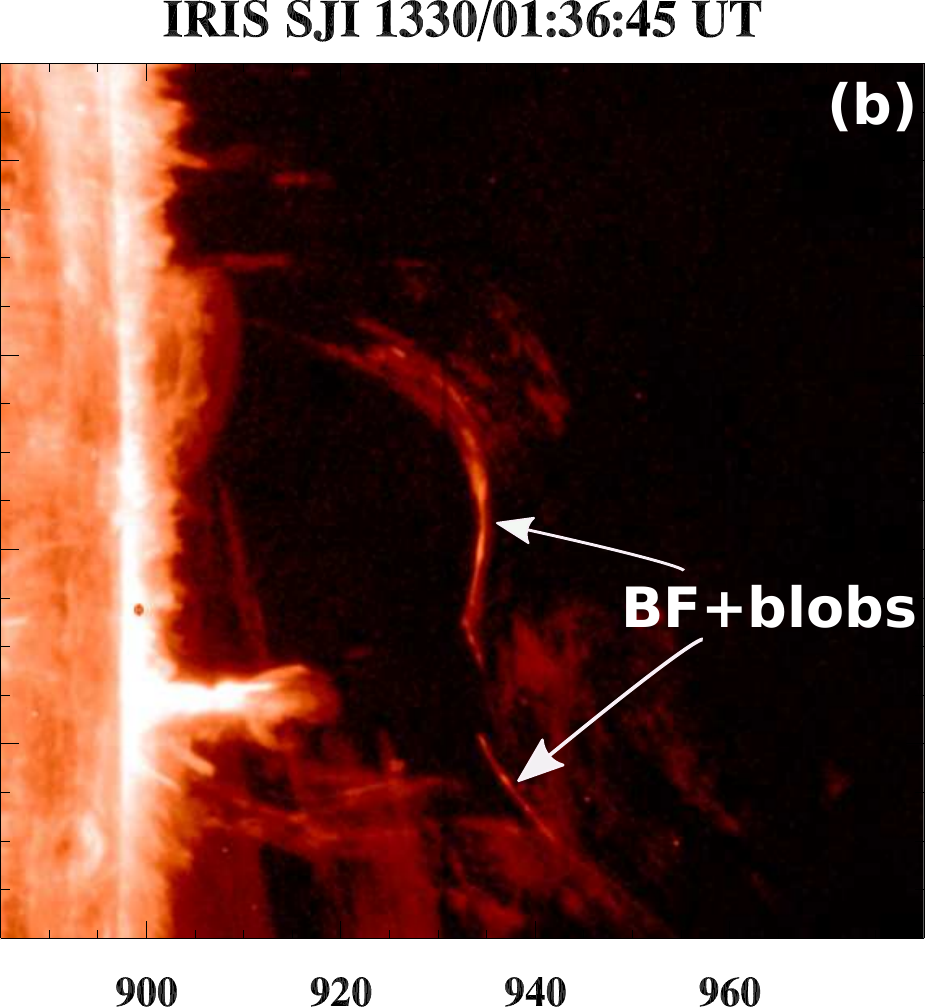}
\includegraphics[width=5.4cm]{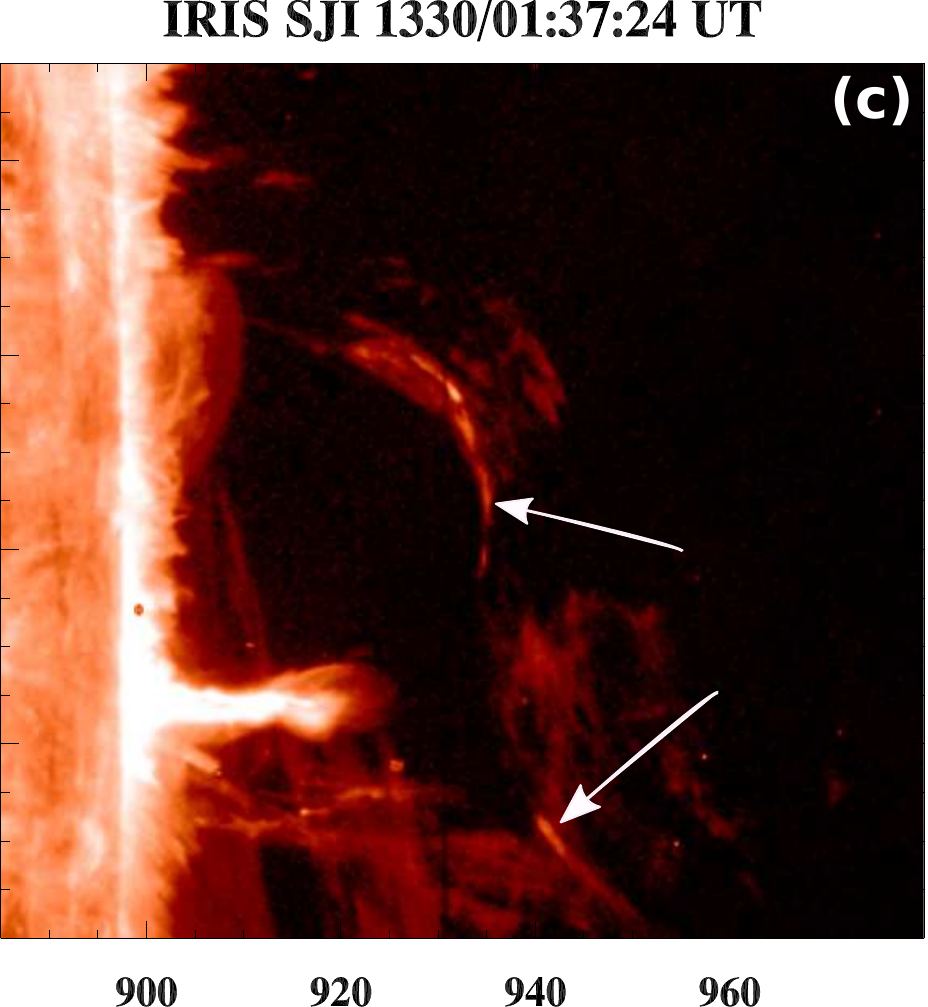}

\includegraphics[width=6.4cm]{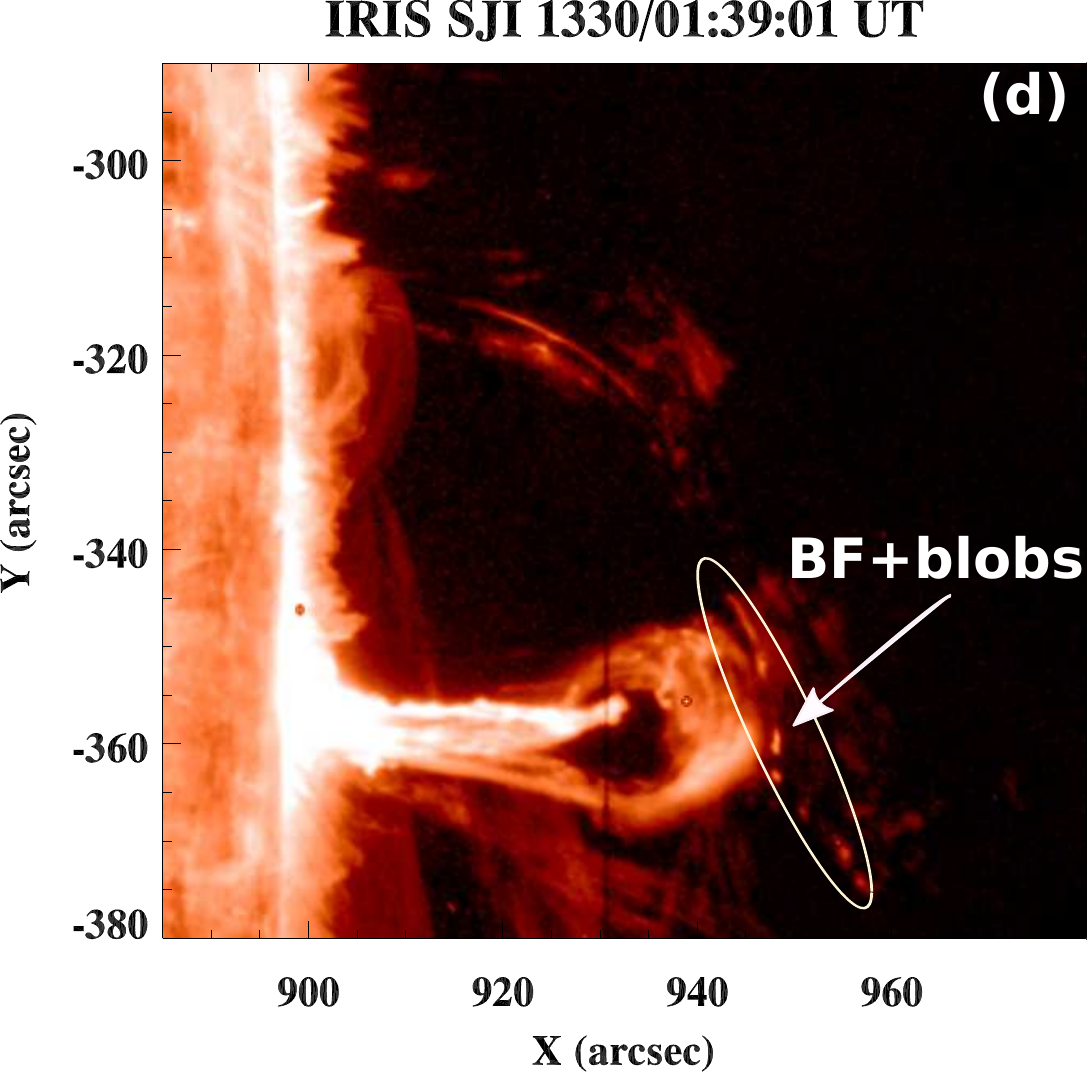}
\includegraphics[width=5.4cm]{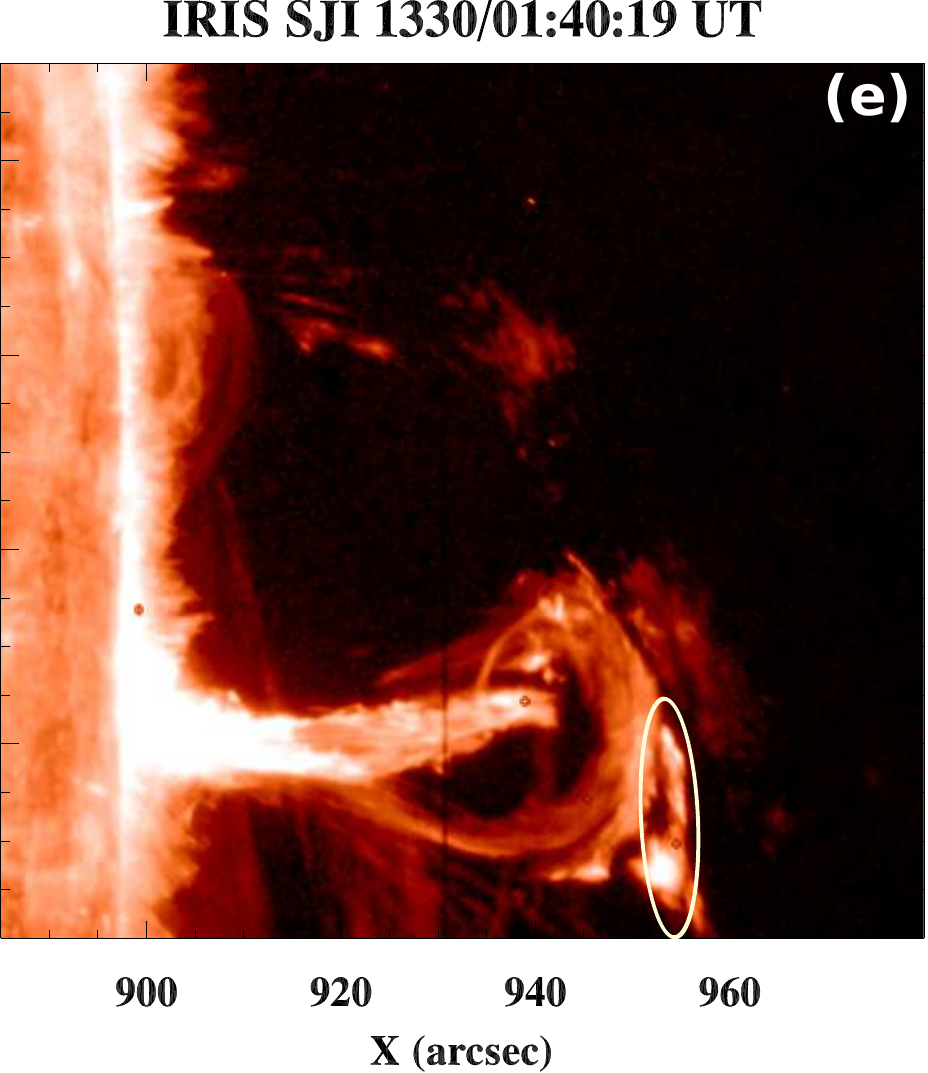}
\includegraphics[width=5.4cm]{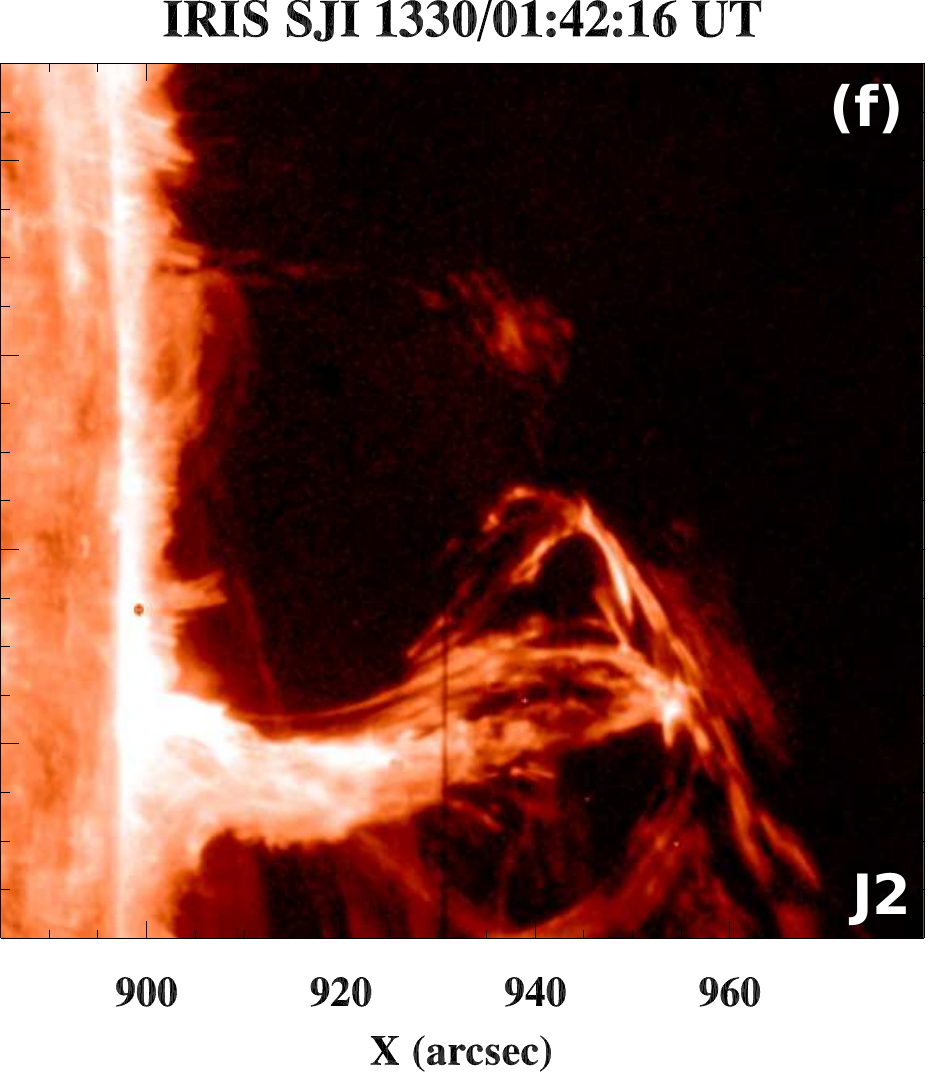}
}
\caption{A sequence of IRIS slit-jaw 1330~\AA\ images of J2. Notations are defined in previous figures.  (An animation of this figure is available online as part of the animation accompanying Figure \ref{fig3}).} 
\label{fig6}
\end{figure*}


\subsection{Jet 2}\label{j2}

After the first jet, the dome structure was illuminated by downflows of jet material as well as re-established coronal rain (Figure \ref{fig3} and accompanying IRIS movie). At 01:30:15 UT, the estimated height of the null point and width of the dome were $\approx$40$\arcsec$ and 60$\arcsec$, respectively. Figure \ref{fig4} and accompanying AIA movies display the initiation and evolution of the second jet (J2) and associated activities in the AIA 304, 171, 193, 131, and 94 \AA\ channels, while Figure \ref{fig5} and accompanying movie cover the same interval in the warm 211~\AA\ channel.  The time-distance plots (Figures \ref{fig5}(g-i)) were made with three slices -- S1 (yellow), S2 (red), and S3 (blue) -- extracted from the AIA 211 \AA\ running difference images (Figure \ref{fig5}(f)). A few minutes after J1 faded out (from 01:08 UT onward), a bright EUV-emitting arc (A) rose slowly above filament F$^\prime$, as shown in Figure \ref{fig5}.  This feature was only detected in the AIA 211 \AA\ running-difference images (Figure \ref{fig5}). Along slice S2 (Figure \ref{fig5}(h)), arc A rose at $v \approx$6 \kms until the start of activity associated with the second jet. Both the IRIS data (accompanying IRIS movie) and the 211 \AA\ time-distance plot show that the filament F$^\prime$ began to rise around 01:30 UT, at $v\approx$7.5 \kms. Thus, the bright arc A and the filament F$^\prime$ rose nearly synchronously from $\approx$01:30 until 01:35 UT (see Figure \ref{fig5}(h)). 

F$^\prime$ and A rose more rapidly after $\approx$01:36 UT, reaching speeds of 80$\pm$13 \kms and 125$\pm$6 \kms, respectively. At the same time, narrow, curved, bright features were observed north and south of the null in all AIA channels (Figure \ref{fig4}(a1-a2,b1-b2,c1-c2)).  These features lengthened with time and propagated bidirectionally at projected speeds on the order of 215$\pm$27 \kms (Figure \ref{fig5}(g)). Although the southward-directed feature was less intense and shorter than its northward counterpart (Figure \ref{fig4}(b1-b2)), both features contained several dynamic bright blobs. During the same interval, some bright loops along the south side of the fan surface separated from the fan and moved southward (see AIA 171~\AA\ movie accompanying Figure \ref{fig3}). In addition, filament F$^\prime$ and its immediate surroundings began to brighten and twist, forming a bright quasi-circular feature (CF) that extended well above the cold, dense filament (see AIA movies accompanying Figure \ref{fig3}).  In addition to the filament F$^\prime$ rooted at its base, CF contained a mixture of cold (absorbing) and hot plasma, as is most clearly evident in the AIA intensity movie at 01:38-01:39 UT. A weak, thin jet also traveled along the spine; until 01:38 or so this flow was visible in all channels, but only appeared in the hottest AIA and Hinode/XRT images thereafter.

As were observed during J1, bidirectional hot plasma flows in the AIA 94 \AA\ images originated near the null when the circular feature encountered the null (Figure \ref{fig4}(e,f)). The plasma blobs are clearer in the cooler channels, suggesting a temperature difference between the blobs and the flows in which they were embedded. When the leading edge of the CF reached the null at 01:39:13 UT, a bright curved feature reappeared above the CF along with multiple bright blobs. The indentation that previously indicated the location of the null was replaced by this feature. A thin dark layer separated the CF from the overlying curved feature, as they continued to rise until 01:40:37 UT.  Subsequently the top of CF and the overlying layers flattened out and became broader, with the distinctions among the layers disappearing around 01:42:13 UT. We also observed apparent coalescence and heating of the plasma blobs in BCS, as they traveled toward the footpoints along the converging field and when the BCS was compressed by arc A or by CF. 

The circular feature opened up at $\approx$01:42:25 UT, when the second jet was launched (Figure \ref{fig4}(f)). Three distinct blobs within J2 along slit S3 (Figure \ref{fig5}(i)) reached projected speeds of $\approx$443$\pm$34, 415$\pm$34, and 246$\pm$69 \kms. Coincident with the J2 onset, three downward-moving blobs on the northern fan loops attained speeds of 215 $\pm$27, 162$\pm$17 and 144$\pm$2 \kms.  Subsequent intermittent upflows and downflows, identified in the running-difference time-intensity plots (Figure \ref{fig5}(g)-(i)), emanated from a site that moved steadily southward and slightly higher along S2, at slower speeds than the initial eruption.  Overall, the second jet was much more complex and explosive than J1, involving the eruption of filament F$^\prime$ and the formation of a bright flare arcade beneath it.  J2 also was much wider than J1, and its main direction of motion shifted steadily southward and eastward, as did the outer spine during the first few minutes of J2.  The jet material first reached the other end of the spine around 01:45 UT, as shown in the AIA 94~\AA\ base and running difference movie. 

A bright linear feature (LF) above the flare arcade appeared around 01:46 UT, then disappeared rapidly after 01:50:25 UT, in all IRIS and AIA channels. Close inspection of the AIA movie reveals bidirectional motions of small blobs in LF at around 01:50 UT.  We extracted the AIA 131 \AA\ relative intensity from a box surrounding the flare arcade (FA in Figures \ref{fig2}(e) and \ref{fig4}((b5,c5)) as a proxy for the flare intensity.  As shown in Figure \ref{fig5}(g), the flare intensity increased swiftly during the rapid elevation of features A and CF, and peaked near the onset of the J2 eruption.  In contrast, the 131 \AA\ intensity in this small region increased very slowly between J1 and J2, as the bright arc A rose.  A second peak in the 131 \AA\ intensity profile, at $\approx$01:48 UT, possibly reflects the removal (through motion and heating) of the absorbing filament material that partially obscured the arcade region until that time, combined with the added emission from LF (compare Figures \ref{fig4}(b4) and (b5)).

After the primary impulsive phase of J2, the brightest emission came from two regions: low loops south of the initial flare arcade but still beneath the fan surface, and long closed loops (outside the fan) beneath the outer spine. These features are not relevant to the primary focus of this paper, which is the breakout reconnection and associated fine structure, but we briefly describe them as they are informative about the post-eruption energetics and relaxation of the AR. Several minutes after J2 onset, a broad region (30-40 arcsec) south of the FCS lit up in the AIA 304 \AA\ and IRIS images, while the same region in the hotter AIA channels was occupied by a mixture of twisting and draining cool and hot material. Irregularly spaced 94~\AA\ bright spots also appeared along the limb south of the flare arcade, at the base of the loops visible in the cooler AIA and IRIS channels. By 02:02 UT, these loops were faintly visible in the coolest emissions while the coronal-temperature images primarily showed some filamentary absorption features rotating against a diffuse glow. 

The extended loop emission is best seen in the AIA 94 \AA\ base-difference movie (left panel) and XRT images, from $\approx$01:47 to the end at 02:42 UT. The loops are brightest at their tops. The bright limb emission sites noted above also persist throughout this period, and appear to extend slowly southward with time.

\section{Discussion}\label{disc}
Two jets of markedly different character occurred in AR 12044 on 2014 May 1. The high-quality observations by SDO and IRIS, combined with our insights derived from modeling coronal jets and CMEs, have enabled us to draw definitive conclusions about the physical mechanisms operating in both dynamic episodes.  Our basic inference from the pre-event observations is that the AR magnetic structure is a fan-spine topology, with a quasi-circular PIL surrounding the minority polarity concentration. 

The first jet did not involve a filament eruption, although F$^\prime$ moved slightly upward and then receded during the jet lifetime. The source of the jet plasma was most likely the corona near the magnetic null, because no cool chromospheric plasma was ejected with the hot jet. Some evidence of energy storage and release associated with the filament channel comes from the EUV observations of a transient bright rim around F$^\prime$ and nearby surface brightenings. No flare arcade was detected, however, which indicates minimal or no flare reconnection below the filament and thus no flux-rope formation along the PIL.  We also interpret the brief appearance of bright spots at the base of the fan, far from the site where flare ribbons would occur, as evidence for particle acceleration on selected field lines comprising the fan surface \citep{sterling2001,kumar2018}. Finally, the bidirectional flows and plasma blobs (plasmoids) coming from the vicinity of the fan-spine intersection are unmistakable signs of magnetic reconnection. 

Therefore, reconnection at the fan-surface current sheet --- that is, breakout reconnection --- is the most compelling explanation for the observed fast outflows and plasmoids moving along the fan surface and spine.  What could have driven this reconnection?  J1 underwent helical motions in the counterclockwise direction, implying that the driving energy came from a twisted magnetic field.  Because we witnessed no signs of flare reconnection in this event, we interpret J1 to be a resistive-kink jet that released only a modest amount of energy and altered the magnetic configuration only slightly.    

Shortly after the first jet, we observed a significant change in behavior of the filament F$^\prime$. The initially faint arc A appeared ahead of F$^\prime$ and drifted upward, then localized EUV brightenings appeared beneath it as the filament rose slowly. The cool filamentary plasma clearly became entwined with hot EUV-emitting plasma as the rise phase progressed. In addition, a growing flare arcade became visible at the base of F$^\prime$, although it was largely obscured by the filament until shortly before J2 onset.  We interpret these phenomena as evidence for the formation and growth of a flux rope enveloping F$^\prime$, with the arc representing the leading edge of the compressed/heated, largely untwisted field around the flux rope. The observed encounter between the rising arc and the constraining fan surface was marked by enhanced emission from the plasma sheet encompassing the BCS, as well as fast bidirectional flows and plasmoids originating at the intersection site.  Shortly before the flows and plasmoids appeared, the bright arc and the filament both accelerated upward. We conclude from this sequence of events that the outer rim of the flux rope began to reconnect with the external field, removing some of the strapping field. The resulting positive feedback accelerated the pace of both flare and breakout reconnection. 

The second jet was generated when the circular feature CF, which trailed the arc A, encountered the fan surface at the BCS. Note that the BCS was significantly lengthened during this phase, as CF pushed against and distorted the fan surface, yielding more opportunities for plasmoids to form. The filament no longer obscured the flare emission, and the arcade approached peak intensity after J2 was initiated, as expected. We interpret CF as the magnetic flux rope formed by flare reconnection beneath F$^\prime$. The resemblance between the circular feature/flux rope and a miniature CME is particularly striking in the AIA 211 \AA\ running-difference images, which is consistent with the universal breakout model. During this energy-release phase, the spine location clearly shifted southward as flux opened in the southern portion of the AR and closed in the northern. As we inferred for J1 and the arc-BCS encounter described above, the fast outflows and embedded plasmoids in the BCS were consequences of breakout reconnection. In contrast with J1, however, J2 involved a filament eruption, although its mass and magnetic field did not escape from the Sun because the flow was directed southward along the closed outer spine. CF rapidly lost its identity as a coherent feature once the eruption started, strongly suggesting that the flux rope released its twist onto external field lines and thus drove the helical jet. 

The bright linear feature (LF) that appeared above the flare arcade at J2 onset disappeared abruptly several minutes before the outflows ceased. LF contained bidirectional plasma flows with plasma blobs; its upper terminus appears to have coincided with the back of CF, although it is difficult to identify CF once the jet began.  We identify LF as the flare current sheet (FCS), highlighted by the dense, heated plasma surrounding it.  Since flare reconnection clearly began many minutes before LF became visible, some combination of the view angle and the structure of the erupting flux rope may have prevented the FCS from being observable until later. In contrast to some observations of flare current sheets persisting for hours behind CMEs \citep{warren2018}, the FCS behind J2 disappeared abruptly within 5-6 minutes. This disappearance is most likely due to the disintegration of the flux rope as a result of breakout reconnection with the external magnetic field. The flux tubes hosting the FCS acquire connections to much more remote footpoints of the field lines near the null, and the plasma can flow freely out of the FCS to fill these newly elongated flux tubes.

The persistent hot (X-ray and EUV) emission from the loops south of the spine has intriguing implications. Either energetic particles remained trapped in these loops for an extended time, or heating by reconnection-driven current filamentation \citep{karpen1996,wyper2016b}, could account for this behavior.

AIA, XRT, and IRIS spectroscopic observations of J2 were studied by \citet{reeves2015}. However, they did not analyze the first jet, and confined much of their attention to the flare reconnection below the filament during the second jet.  We agree with them on the role of flare reconnection in forming the rising flux rope in which the filament was embedded, but disagree with their conclusion that J2 was triggered by tether-cutting. Our analysis finds ample evidence for the crucial role of breakout reconnection in J1, and in opening up the flux rope and releasing both cool filament and hot coronal plasma in J2. 

\section{Conclusions}\label{conc}
We present here the first evidence of plasmoid formation during breakout reconnection in two successive coronal jets using AIA and IRIS observations. The key new feature of our results is that they demonstrate the existence and importance of plasmoids during fast reconnection in a fully 3D magnetic configuration. Plasmoids have long been observed in both simulations and observations of reconnection in 2D or quasi-2D (i.e., flare) current sheets, but they have never been observed in a fully 3D current sheet on the Sun. MHD simulations of the interaction between emerging flux and pre-existing coronal flux have produced plasmoids as well \citep{yokoyama1995,archontis2006,moreno2013}. However, no signs of flux emergence were observed prior to J1 or J2. Both were limb events for which magnetograms were not available, but flux emergence would have caused the coronal magnetic field to expand and change connectivity to form the fan-spine topology ``anemone"; instead, the configuration appeared quite stable and unchanged for many hours prior to jet onset. Our prior analysis of 27 jets in an equatorial coronal hole found no direct correlation between emerging flux and the observed activity: the bright point sources emerged hours to days before any jets were generated \citep{kumar2019}. Our present observations reveal that plasmoids form in the 3D BCS, which is consistent with recent high-resolution 3D MHD simulations of breakout reconnection \citep{edmondson2010,wyper2014a,wyper2014b,edmondson2017}. Our new observations imply that, in fact, plasmoids are a central feature of all fast reconnection.

The first jet is consistent with the predictions of our resistive-kink model \citep{pariat2009}, while the second jet agrees with our breakout model for solar eruptions \citep{wyper2017}.  The properties of the blobs (size, temperature, etc.) were similar to those reported for the plasma blobs in previous AIA observations of flare current sheets \citep{takasao2012,kumar2013,kumar2018,kumar2019}. These jet observations also agree with results of the highest-resolution MHD simulations of breakout jets and CMEs, which consistently find that multiple plasmoids form in the breakout and flare current sheets \citep{karpen2012,guo2013,lynch2013,lynch2016,guidoni2016}. Additional observations with multiwavelength data sets (EUV, X-ray, and radio) of breakout current sheets and plasmoids will be extremely important for testing current models of eruption mechanisms, and ultimately for understanding the initiation of the broad range of solar eruptions (mini-CMEs and jets to large CMEs) and particle acceleration in the BCS in these events. Continuity, high spatial resolution, and rapid cadence of observations are particularly crucial for such studies, as many of the key reconnection signatures are transient and small scale. 

The combination of a resistive-kink jet followed by a more powerful breakout jet deduced from these observations also is novel, and demonstrates that the characteristics of energy buildup in the filament channel play a key role in determining both the morphology and the energetics of eruptions.  For limb events, unfortunately, it is impossible to determine the magnetic configuration accurately within less than a few days before and after the time of eruption. Multiple magnetographs and EUV imagers distributed around the Sun would aid enormously in deciphering the pre-eruptive structure and its evolution throughout an eruption. 

\acknowledgments
 SDO is a mission for NASA's Living With a Star (LWS) program. IRIS is a NASA small Explorer mission developed and operated by LMSAL with mission operations executed at NASA Ames Research center and major contributions to downlink communications funded by ESA and the Norwegian Space Centre. Hinode is a Japanese mission developed and launched by ISAS/JAXA, with NAOJ as domestic partner and NASA and STFC (UK) as international partners. It is operated by these agencies in co-operation with ESA and NSC (Norway). This research was supported by grants from the NASA H-SR and H-ISFM programs. and through P.F.W.'s award of a Royal Astronomical Society Fellowship. The magnetic-field extrapolation and simulation visualizations were produced with VAPOR (www.vapor.ucar.edu), a product of the Computational Information Systems Laboratory at the National Center for Atmospheric Research.


\bibliographystyle{aasjournal}
\bibliography{reference.bib}

\begin{thebibliography}{}
\expandafter\ifx\csname natexlab\endcsname\relax\def\natexlab#1{#1}\fi
\providecommand{\url}[1]{\href{#1}{#1}}
\providecommand{\dodoi}[1]{doi:~\href{http://doi.org/#1}{\nolinkurl{#1}}}
\providecommand{\doeprint}[1]{\href{http://ascl.net/#1}{\nolinkurl{http://ascl.net/#1}}}
\providecommand{\doarXiv}[1]{\href{https://arxiv.org/abs/#1}{\nolinkurl{https://arxiv.org/abs/#1}}}

\bibitem[{{Antiochos}(1998)}]{antiochos1998}
{Antiochos}, S.~K. 1998, ApJL, 502, L181, \dodoi{10.1086/311507}

\bibitem[{{Antiochos} {et~al.}(1999){Antiochos}, {DeVore}, \&
  {Klimchuk}}]{antiochos1999}
{Antiochos}, S.~K., {DeVore}, C.~R., \& {Klimchuk}, J.~A. 1999, ApJ, 510, 485,
  \dodoi{10.1086/306563}

\bibitem[{{Archontis} {et~al.}(2006){Archontis}, {Galsgaard},
  {Moreno-Insertis}, \& {Hood}}]{archontis2006}
{Archontis}, V., {Galsgaard}, K., {Moreno-Insertis}, F., \& {Hood}, A.~W. 2006,
  \apjl, 645, L161, \dodoi{10.1086/506203}

\bibitem[{{Aurass} {et~al.}(2013){Aurass}, {Holman}, {Braune}, {Mann}, \&
  {Zlobec}}]{aurass2013}
{Aurass}, H., {Holman}, G., {Braune}, S., {Mann}, G., \& {Zlobec}, P. 2013,
  Astronomy and Astrophysics, 555, A40, \dodoi{10.1051/0004-6361/201321111}

\bibitem[{{Bhattacharjee} {et~al.}(2009){Bhattacharjee}, {Huang}, {Yang}, \&
  {Rogers}}]{bhattacharjee2009}
{Bhattacharjee}, A., {Huang}, Y.-M., {Yang}, H., \& {Rogers}, B. 2009, Physics
  of Plasmas, 16, 112102, \dodoi{10.1063/1.3264103}

\bibitem[{{De Pontieu} {et~al.}(2014){De Pontieu}, {Title}, {Lemen}, {Kushner},
  {Akin}, {Allard}, {Berger}, {Boerner}, {Cheung}, {Chou}, {Drake}, {Duncan},
  {Freeland}, {Heyman}, {Hoffman}, {Hurlburt}, {Lindgren}, {Mathur}, {Rehse},
  {Sabolish}, {Seguin}, {Schrijver}, {Tarbell}, {W{\"u}lser}, {Wolfson},
  {Yanari}, {Mudge}, {Nguyen-Phuc}, {Timmons}, {van Bezooijen}, {Weingrod},
  {Brookner}, {Butcher}, {Dougherty}, {Eder}, {Knagenhjelm}, {Larsen},
  {Mansir}, {Phan}, {Boyle}, {Cheimets}, {DeLuca}, {Golub}, {Gates}, {Hertz},
  {McKillop}, {Park}, {Perry}, {Podgorski}, {Reeves}, {Saar}, {Testa}, {Tian},
  {Weber}, {Dunn}, {Eccles}, {Jaeggli}, {Kankelborg}, {Mashburn}, {Pust},
  {Springer}, {Carvalho}, {Kleint}, {Marmie}, {Mazmanian}, {Pereira}, {Sawyer},
  {Strong}, {Worden}, {Carlsson}, {Hansteen}, {Leenaarts}, {Wiesmann},
  {Aloise}, {Chu}, {Bush}, {Scherrer}, {Brekke}, {Martinez-Sykora}, {Lites},
  {McIntosh}, {Uitenbroek}, {Okamoto}, {Gummin}, {Auker}, {Jerram}, {Pool}, \&
  {Waltham}}]{de-pontieu2014}
{De Pontieu}, B., {Title}, A.~M., {Lemen}, J.~R., {et~al.} 2014, Solar Physics,
  289, 2733, \dodoi{10.1007/s11207-014-0485-y}

\bibitem[{{DeForest}(2017)}]{deforest2017}
{DeForest}, C.~E. 2017, ApJ, 838, 155, \dodoi{10.3847/1538-4357/aa67f1}

\bibitem[{{DeVore} \& {Antiochos}(2008)}]{devore2008}
{DeVore}, C.~R., \& {Antiochos}, S.~K. 2008, ApJ, 680, 740,
  \dodoi{10.1086/588011}

\bibitem[{{Drake} {et~al.}(2006){Drake}, {Swisdak}, {Che}, \&
  {Shay}}]{drake2006}
{Drake}, J.~F., {Swisdak}, M., {Che}, H., \& {Shay}, M.~A. 2006, Nature, 443,
  553, \dodoi{10.1038/nature05116}

\bibitem[{{Edmondson} {et~al.}(2010){Edmondson}, {Antiochos}, {DeVore}, \&
  {Zurbuchen}}]{edmondson2010}
{Edmondson}, J.~K., {Antiochos}, S.~K., {DeVore}, C.~R., \& {Zurbuchen}, T.~H.
  2010, \apj, 718, 72, \dodoi{10.1088/0004-637X/718/1/72}

\bibitem[{{Edmondson} \& {Lynch}(2017)}]{edmondson2017}
{Edmondson}, J.~K., \& {Lynch}, B.~J. 2017, \apj, 849, 28,
  \dodoi{10.3847/1538-4357/aa83ba}

\bibitem[{{Fletcher} {et~al.}(2001){Fletcher}, {Metcalf}, {Alexander}, {Brown},
  \& {Ryder}}]{fletcher2001}
{Fletcher}, L., {Metcalf}, T.~R., {Alexander}, D., {Brown}, D.~S., \& {Ryder},
  L.~A. 2001, ApJ, 554, 451, \dodoi{10.1086/321377}

\bibitem[{{Forbes} \& {Acton}(1996)}]{forbes1996}
{Forbes}, T.~G., \& {Acton}, L.~W. 1996, ApJ, 459, 330, \dodoi{10.1086/176896}

\bibitem[{{Gary} {et~al.}(2018){Gary}, {Chen}, {Dennis}, {Fleishman},
  {Hurford}, {Krucker}, {McTiernan}, {Nita}, {Shih}, {White}, \&
  {Yu}}]{gary2018}
{Gary}, D.~E., {Chen}, B., {Dennis}, B.~R., {et~al.} 2018, ApJ, 863, 83,
  \dodoi{10.3847/1538-4357/aad0ef}

\bibitem[{{Golub} {et~al.}(2007){Golub}, {Deluca}, {Austin}, {Bookbinder},
  {Caldwell}, {Cheimets}, {Cirtain}, {Cosmo}, {Reid}, {Sette}, {Weber},
  {Sakao}, {Kano}, {Shibasaki}, {Hara}, {Tsuneta}, {Kumagai}, {Tamura},
  {Shimojo}, {McCracken}, {Carpenter}, {Haight}, {Siler}, {Wright}, {Tucker},
  {Rutledge}, {Barbera}, {Peres}, \& {Varisco}}]{golub2007}
{Golub}, L., {Deluca}, E., {Austin}, G., {et~al.} 2007, Solar Physics, 243, 63,
  \dodoi{10.1007/s11207-007-0182-1}

\bibitem[{{Guidoni} {et~al.}(2016){Guidoni}, {DeVore}, {Karpen}, \&
  {Lynch}}]{guidoni2016}
{Guidoni}, S.~E., {DeVore}, C.~R., {Karpen}, J.~T., \& {Lynch}, B.~J. 2016,
  ApJ, 820, 60, \dodoi{10.3847/0004-637X/820/1/60}

\bibitem[{{Guo} {et~al.}(2013){Guo}, {Bhattacharjee}, \& {Huang}}]{guo2013}
{Guo}, L.-J., {Bhattacharjee}, A., \& {Huang}, Y.-M. 2013, ApJL, 771, L14,
  \dodoi{10.1088/2041-8205/771/1/L14}

\bibitem[{{Huang} {et~al.}(2013){Huang}, {Bhattacharjee}, \&
  {Forbes}}]{huang2013}
{Huang}, Y.-M., {Bhattacharjee}, A., \& {Forbes}, T.~G. 2013, Physics of
  Plasmas, 20, 082131, \dodoi{10.1063/1.4819715}

\bibitem[{{Innes} {et~al.}(2016){Innes}, {Bu{\v c}{\'{\i}}k}, {Guo}, \&
  {Nitta}}]{innes2016}
{Innes}, D.~E., {Bu{\v c}{\'{\i}}k}, R., {Guo}, L.-J., \& {Nitta}, N. 2016,
  Astronomische Nachrichten, 337, 1024, \dodoi{10.1002/asna.201612428}

\bibitem[{Karpen {et~al.}(2019)Karpen, Kumar, Antiochos, Gary, \&
  Dahlin}]{karpen2019}
Karpen, J., Kumar, P., Antiochos, S., Gary, D., \& Dahlin, J. 2019, ApJ, in
  preparation

\bibitem[{{Karpen} {et~al.}(1996){Karpen}, {Antiochos}, \&
  {DeVore}}]{karpen1996}
{Karpen}, J.~T., {Antiochos}, S.~K., \& {DeVore}, C.~R. 1996, ApJL, 460, L73,
  \dodoi{10.1086/309965}

\bibitem[{{Karpen} {et~al.}(2012){Karpen}, {Antiochos}, \&
  {DeVore}}]{karpen2012}
---. 2012, ApJ, 760, 81, \dodoi{10.1088/0004-637X/760/1/81}

\bibitem[{{Karpen} {et~al.}(2017){Karpen}, {DeVore}, {Antiochos}, \&
  {Pariat}}]{karpen2017}
{Karpen}, J.~T., {DeVore}, C.~R., {Antiochos}, S.~K., \& {Pariat}, E. 2017,
  ApJ, 834, 62, \dodoi{10.3847/1538-4357/834/1/62}

\bibitem[{{Kumar} \& {Cho}(2013)}]{kumar2013}
{Kumar}, P., \& {Cho}, K.-S. 2013, Astronomy and Astrophysics, 557, A115,
  \dodoi{10.1051/0004-6361/201220999}

\bibitem[{{Kumar} \& {Innes}(2013)}]{kumar2013a}
{Kumar}, P., \& {Innes}, D.~E. 2013, Solar Physics, 288, 255,
  \dodoi{10.1007/s11207-013-0303-y}

\bibitem[{{Kumar} {et~al.}(2016){Kumar}, {Innes}, \& {Cho}}]{kumar2016}
{Kumar}, P., {Innes}, D.~E., \& {Cho}, K.-S. 2016, ApJ, 828, 28,
  \dodoi{10.3847/0004-637X/828/1/28}

\bibitem[{{Kumar} {et~al.}(2018){Kumar}, {Karpen}, {Antiochos}, {Wyper},
  {DeVore}, \& {DeForest}}]{kumar2018}
{Kumar}, P., {Karpen}, J.~T., {Antiochos}, S.~K., {et~al.} 2018, ApJ, 854, 155,
  \dodoi{10.3847/1538-4357/aaab4f}

\bibitem[{{Kumar} {et~al.}(2019){Kumar}, {Karpen}, {Antiochos}, {Wyper},
  {DeVore}, \& {DeForest}}]{kumar2019}
---. 2019, ApJ, 873, 93, \dodoi{10.3847/1538-4357/ab04af}

\bibitem[{{Kumar} {et~al.}(2017){Kumar}, {Nakariakov}, \& {Cho}}]{kumar2017}
{Kumar}, P., {Nakariakov}, V.~M., \& {Cho}, K.-S. 2017, ApJ, 844, 149,
  \dodoi{10.3847/1538-4357/aa7d53}

\bibitem[{{Kumar} {et~al.}(2015){Kumar}, {Yurchyshyn}, {Wang}, \&
  {Cho}}]{kumar2015}
{Kumar}, P., {Yurchyshyn}, V., {Wang}, H., \& {Cho}, K.-S. 2015, ApJ, 809, 83,
  \dodoi{10.1088/0004-637X/809/1/83}

\bibitem[{{Lemen} {et~al.}(2012){Lemen}, {Title}, {Akin}, {Boerner}, {Chou},
  {Drake}, {Duncan}, {Edwards}, {Friedlaender}, {Heyman}, {Hurlburt}, {Katz},
  {Kushner}, {Levay}, {Lindgren}, {Mathur}, {McFeaters}, {Mitchell}, {Rehse},
  {Schrijver}, {Springer}, {Stern}, {Tarbell}, {Wuelser}, {Wolfson}, {Yanari},
  {Bookbinder}, {Cheimets}, {Caldwell}, {Deluca}, {Gates}, {Golub}, {Park},
  {Podgorski}, {Bush}, {Scherrer}, {Gummin}, {Smith}, {Auker}, {Jerram},
  {Pool}, {Soufli}, {Windt}, {Beardsley}, {Clapp}, {Lang}, \&
  {Waltham}}]{lemen2012}
{Lemen}, J.~R., {Title}, A.~M., {Akin}, D.~J., {et~al.} 2012, Solar Physics,
  275, 17, \dodoi{10.1007/s11207-011-9776-8}

\bibitem[{{Li} \& {Yang}(2019)}]{li2019}
{Li}, H., \& {Yang}, J. 2019, ApJ, 872, 87, \dodoi{10.3847/1538-4357/aafb3a}

\bibitem[{{Loureiro} {et~al.}(2007){Loureiro}, {Schekochihin}, \&
  {Cowley}}]{loureiro2007}
{Loureiro}, N.~F., {Schekochihin}, A.~A., \& {Cowley}, S.~C. 2007, Physics of
  Plasmas, 14, 100703, \dodoi{10.1063/1.2783986}

\bibitem[{{Lynch} {et~al.}(2008){Lynch}, {Antiochos}, {DeVore}, {Luhmann}, \&
  {Zurbuchen}}]{lynch2008}
{Lynch}, B.~J., {Antiochos}, S.~K., {DeVore}, C.~R., {Luhmann}, J.~G., \&
  {Zurbuchen}, T.~H. 2008, ApJ, 683, 1192, \dodoi{10.1086/589738}

\bibitem[{{Lynch} {et~al.}(2009){Lynch}, {Antiochos}, {Li}, {Luhmann}, \&
  {DeVore}}]{lynch2009}
{Lynch}, B.~J., {Antiochos}, S.~K., {Li}, Y., {Luhmann}, J.~G., \& {DeVore},
  C.~R. 2009, ApJ, 697, 1918, \dodoi{10.1088/0004-637X/697/2/1918}

\bibitem[{{Lynch} \& {Edmondson}(2013)}]{lynch2013}
{Lynch}, B.~J., \& {Edmondson}, J.~K. 2013, ApJ, 764, 87,
  \dodoi{10.1088/0004-637X/764/1/87}

\bibitem[{{Lynch} {et~al.}(2016){Lynch}, {Edmondson}, {Kazachenko}, \&
  {Guidoni}}]{lynch2016}
{Lynch}, B.~J., {Edmondson}, J.~K., {Kazachenko}, M.~D., \& {Guidoni}, S.~E.
  2016, ApJ, 826, 43, \dodoi{10.3847/0004-637X/826/1/43}

\bibitem[{{MacNeice} {et~al.}(2004){MacNeice}, {Antiochos}, {Phillips},
  {Spicer}, {DeVore}, \& {Olson}}]{macneice2004}
{MacNeice}, P., {Antiochos}, S.~K., {Phillips}, A., {et~al.} 2004, ApJ, 614,
  1028, \dodoi{10.1086/423887}

\bibitem[{{Mason} {et~al.}(2019){Mason}, {Antiochos}, \& {Viall}}]{mason2019}
{Mason}, E.~I., {Antiochos}, S.~K., \& {Viall}, N.~M. 2019, ApJL, 874, L33,
  \dodoi{10.3847/2041-8213/ab0c5d}

\bibitem[{{Masson} {et~al.}(2009){Masson}, {Pariat}, {Aulanier}, \&
  {Schrijver}}]{masson2009}
{Masson}, S., {Pariat}, E., {Aulanier}, G., \& {Schrijver}, C.~J. 2009, ApJ,
  700, 559, \dodoi{10.1088/0004-637X/700/1/559}

\bibitem[{{McGlasson} {et~al.}(2019){McGlasson}, {Panesar}, {Sterling}, \&
  {Moore}}]{McGlasson2019}
{McGlasson}, R.~A., {Panesar}, N.~K., {Sterling}, A.~C., \& {Moore}, R. 2019,
  arXiv e-prints.
\newblock \doarXiv{1906.06452}

\bibitem[{{Moore} {et~al.}(2010){Moore}, {Cirtain}, {Sterling}, \&
  {Falconer}}]{moore2010}
{Moore}, R.~L., {Cirtain}, J.~W., {Sterling}, A.~C., \& {Falconer}, D.~A. 2010,
  ApJ, 720, 757, \dodoi{10.1088/0004-637X/720/1/757}

\bibitem[{{Moore} {et~al.}(2018){Moore}, {Sterling}, \& {Panesar}}]{moore2018}
{Moore}, R.~L., {Sterling}, A.~C., \& {Panesar}, N.~K. 2018, \apj, 859, 3,
  \dodoi{10.3847/1538-4357/aabe79}

\bibitem[{{Moreno-Insertis} \& {Galsgaard}(2013)}]{moreno2013}
{Moreno-Insertis}, F., \& {Galsgaard}, K. 2013, ApJ, 771, 20,
  \dodoi{10.1088/0004-637X/771/1/20}

\bibitem[{{Panesar} {et~al.}(2018){Panesar}, {Sterling}, \&
  {Moore}}]{panesar2018}
{Panesar}, N.~K., {Sterling}, A.~C., \& {Moore}, R.~L. 2018, \apj, 853, 189,
  \dodoi{10.3847/1538-4357/aaa3e9}

\bibitem[{{Pariat} {et~al.}(2009){Pariat}, {Antiochos}, \&
  {DeVore}}]{pariat2009}
{Pariat}, E., {Antiochos}, S.~K., \& {DeVore}, C.~R. 2009, ApJ, 691, 61,
  \dodoi{10.1088/0004-637X/691/1/61}

\bibitem[{{Pariat} {et~al.}(2010){Pariat}, {Antiochos}, \&
  {DeVore}}]{pariat2010}
---. 2010, ApJ, 714, 1762, \dodoi{10.1088/0004-637X/714/2/1762}

\bibitem[{{Pariat} {et~al.}(2015){Pariat}, {Dalmasse}, {DeVore}, {Antiochos},
  \& {Karpen}}]{pariat2015}
{Pariat}, E., {Dalmasse}, K., {DeVore}, C.~R., {Antiochos}, S.~K., \& {Karpen},
  J.~T. 2015, Astronomy and Astrophysics, 573, A130,
  \dodoi{10.1051/0004-6361/201424209}

\bibitem[{{Pariat} {et~al.}(2016){Pariat}, {Dalmasse}, {DeVore}, {Antiochos},
  \& {Karpen}}]{pariat2016}
---. 2016, Astronomy and Astrophysics, 596, A36,
  \dodoi{10.1051/0004-6361/201629109}

\bibitem[{{Patsourakos} {et~al.}(2008){Patsourakos}, {Pariat}, {Vourlidas},
  {Antiochos}, \& {Wuelser}}]{patsourakos2008}
{Patsourakos}, S., {Pariat}, E., {Vourlidas}, A., {Antiochos}, S.~K., \&
  {Wuelser}, J.~P. 2008, ApJL, 680, L73, \dodoi{10.1086/589769}

\bibitem[{{Priest} \& {Titov}(1996)}]{priest1996}
{Priest}, E.~R., \& {Titov}, V.~S. 1996, Philosophical Transactions of the
  Royal Society of London Series A, 354, 2951, \dodoi{10.1098/rsta.1996.0136}

\bibitem[{{Raouafi} {et~al.}(2010){Raouafi}, {Georgoulis}, {Rust}, \&
  {Bernasconi}}]{raouafi2010}
{Raouafi}, N.-E., {Georgoulis}, M.~K., {Rust}, D.~M., \& {Bernasconi}, P.~N.
  2010, ApJ, 718, 981, \dodoi{10.1088/0004-637X/718/2/981}

\bibitem[{{Raouafi} {et~al.}(2016){Raouafi}, {Patsourakos}, {Pariat}, {Young},
  {Sterling}, {Savcheva}, {Shimojo}, {Moreno-Insertis}, {DeVore}, {Archontis},
  {T{\"o}r{\"o}k}, {Mason}, {Curdt}, {Meyer}, {Dalmasse}, \&
  {Matsui}}]{raouafi2016}
{Raouafi}, N.~E., {Patsourakos}, S., {Pariat}, E., {et~al.} 2016, Space Sci.
  Rev., 201, 1, \dodoi{10.1007/s11214-016-0260-5}

\bibitem[{{Reeves} {et~al.}(2015){Reeves}, {McCauley}, \& {Tian}}]{reeves2015}
{Reeves}, K.~K., {McCauley}, P.~I., \& {Tian}, H. 2015, ApJ, 807, 7,
  \dodoi{10.1088/0004-637X/807/1/7}

\bibitem[{{Schou} {et~al.}(2012){Schou}, {Scherrer}, {Bush}, {Wachter},
  {Couvidat}, {Rabello-Soares}, {Bogart}, {Hoeksema}, {Liu}, {Duvall}, {Akin},
  {Allard}, {Miles}, {Rairden}, {Shine}, {Tarbell}, {Title}, {Wolfson},
  {Elmore}, {Norton}, \& {Tomczyk}}]{schou2012}
{Schou}, J., {Scherrer}, P.~H., {Bush}, R.~I., {et~al.} 2012, Solar Physics,
  275, 229, \dodoi{10.1007/s11207-011-9842-2}

\bibitem[{{Shibata} {et~al.}(1994){Shibata}, {Nitta}, {Strong}, {Matsumoto},
  {Yokoyama}, {Hirayama}, {Hudson}, \& {Ogawara}}]{shibata1994}
{Shibata}, K., {Nitta}, N., {Strong}, K.~T., {et~al.} 1994, ApJL, 431, L51,
  \dodoi{10.1086/187470}

\bibitem[{{Sterling} \& {Moore}(2001)}]{sterling2001}
{Sterling}, A.~C., \& {Moore}, R.~L. 2001, ApJ, 560, 1045,
  \dodoi{10.1086/322241}

\bibitem[{{Sterling} {et~al.}(2015){Sterling}, {Moore}, {Falconer}, \&
  {Adams}}]{sterling2015}
{Sterling}, A.~C., {Moore}, R.~L., {Falconer}, D.~A., \& {Adams}, M. 2015,
  Nature, 523, 437, \dodoi{10.1038/nature14556}

\bibitem[{{Takasao} {et~al.}(2012){Takasao}, {Asai}, {Isobe}, \&
  {Shibata}}]{takasao2012}
{Takasao}, S., {Asai}, A., {Isobe}, H., \& {Shibata}, K. 2012, ApJL, 745, L6,
  \dodoi{10.1088/2041-8205/745/1/L6}

\bibitem[{{Wang} \& {Liu}(2012)}]{wang2012}
{Wang}, H., \& {Liu}, C. 2012, ApJ, 760, 101,
  \dodoi{10.1088/0004-637X/760/2/101}

\bibitem[{{Warren} {et~al.}(2018){Warren}, {Brooks}, {Ugarte-Urra}, {Reep},
  {Crump}, \& {Doschek}}]{warren2018}
{Warren}, H.~P., {Brooks}, D.~H., {Ugarte-Urra}, I., {et~al.} 2018, ApJ, 854,
  122, \dodoi{10.3847/1538-4357/aaa9b8}

\bibitem[{{Wyper} {et~al.}(2017){Wyper}, {Antiochos}, \& {DeVore}}]{wyper2017}
{Wyper}, P.~F., {Antiochos}, S.~K., \& {DeVore}, C.~R. 2017, Nature, 544, 452,
  \dodoi{10.1038/nature22050}

\bibitem[{{Wyper} \& {DeVore}(2016)}]{wyper2016a}
{Wyper}, P.~F., \& {DeVore}, C.~R. 2016, ApJ, 820, 77,
  \dodoi{10.3847/0004-637X/820/1/77}

\bibitem[{{Wyper} {et~al.}(2018){Wyper}, {DeVore}, \& {Antiochos}}]{wyper2018}
{Wyper}, P.~F., {DeVore}, C.~R., \& {Antiochos}, S.~K. 2018, ApJ, 852, 98,
  \dodoi{10.3847/1538-4357/aa9ffc}

\bibitem[{{Wyper} {et~al.}(2016){Wyper}, {DeVore}, {Karpen}, \&
  {Lynch}}]{wyper2016b}
{Wyper}, P.~F., {DeVore}, C.~R., {Karpen}, J.~T., \& {Lynch}, B.~J. 2016, ApJ,
  827, 4, \dodoi{10.3847/0004-637X/827/1/4}

\bibitem[{{Wyper} \& {Pontin}(2014{\natexlab{a}})}]{wyper2014a}
{Wyper}, P.~F., \& {Pontin}, D.~I. 2014{\natexlab{a}}, Physics of Plasmas, 21,
  082114, \dodoi{10.1063/1.4893149}

\bibitem[{{Wyper} \& {Pontin}(2014{\natexlab{b}})}]{wyper2014b}
---. 2014{\natexlab{b}}, Physics of Plasmas, 21, 102102,
  \dodoi{10.1063/1.4896060}

\bibitem[{{Yokoyama} \& {Shibata}(1995)}]{yokoyama1995}
{Yokoyama}, T., \& {Shibata}, K. 1995, \nat, 375, 42, \dodoi{10.1038/375042a0}

\end{thebibliography}

\end{document}